\documentclass[10pt]{article}
\pdfoutput=1
\usepackage{jheparxiv}
\usepackage[utf8]{inputenc}
\usepackage{amsmath}
\usepackage{amsfonts}
\usepackage{amssymb}
\usepackage{latexsym}
\usepackage{mathrsfs}
\usepackage{graphicx}
\usepackage{color}
\usepackage[all]{xy}
\usepackage[usenames,dvipsnames,svgnames]{xcolor}

\newcommand{\df}{{\rm d}}
\newcommand{\h}{\mathfrak{h}}
\newcommand{\dbar}{\mathchar'26\mkern-12mu d }
\newcommand{\deltabar}{\mathchar'26\mkern-9mu \delta }

\subheader{\hfill \begin{tabular}{r} \texttt{QMUL-PH-19-34} \end{tabular}}
 
\title{\Large The Classical Double Copy of a Point Charge}
\author[a,b]{Kwangeon Kim,}
\author[a]{Kanghoon Lee,}
\author[c]{Ricardo Monteiro,}
\author[d]{Isobel Nicholson,}
\author[c]{David Peinador Veiga}

\affiliation[a]{Center for Theoretical Physics of the Universe, Institute for Basic Science (IBS), \\ 55, Expo-ro, Yuseong-gu, Daejeon 34126, Korea}
\affiliation[b]{Department of Physics, Yonsei University, Seoul 03722, Korea}
\affiliation[c]{Centre for Research in String Theory, School of Physics and Astronomy, \\
        Queen Mary University of London, E1 4NS, United Kingdom}
\affiliation[d]{Higgs Centre for Theoretical Physics, School of Physics and Astronomy, \\
The University of Edinburgh, EH9 3JZ, Scotland, United Kingdom}
        
\abstract{

The classical double copy relates solutions to the equations of motion in gauge theory and in gravity. In this paper, we present two double-copy formalisms for relating the Coulomb solution in gauge theory to the two-parameter Janis-Newman-Winicour solution in gravity. The latter is a static, spherically symmetric, asymptotically flat solution that generically includes a dilaton field, but also admits the Schwarzschild solution as a special case. We first present the classical double copy as a perturbative construction, similar to its formulation for scattering amplitudes, and then present it as an exact map, with a novel generalisation of the Kerr-Schild double copy motivated by double field theory. The latter formalism exhibits the relation between the Kerr-Schild classical double copy and the string theory origin of the double copy for scattering amplitudes.

}

\begin{document}

\maketitle


\section{Introduction}

The {\it double copy} is a relation between gauge theory and gravity, which originated in the structure of scattering amplitudes in these theories. It was first noticed in the context of string theory \cite{Kawai:1985xq}, and was later explored for the calculation of scattering amplitudes in gravity theories \cite{Bern:2008qj,Bern:2010ue}, where it became a formidable tool. See ref.~\cite{Bern:2019prr} for a recent review.

The intuitive idea of the double copy is that calculations for spin-2 particles are somehow `factorisable' into calculations for spin-1 particles. If we think in terms of the representations of the Lorentz group for massless particles, however, we notice imediately that the tensor product of spin-1 states leads to more than spin-2 states. Take the basis of positive and negative helicities $\{\epsilon^+_\mu,\epsilon^-_\mu\}$ for spin-1 states in four dimensions. Then the `product-gravity' states are $\{\epsilon^+_\mu \epsilon^+_\nu,\epsilon^-_\mu \epsilon^-_\nu,\epsilon^+_\mu \epsilon^-_\nu,\epsilon^-_\mu \epsilon^+_\nu\}$, the first two of which are graviton states of positive and negative helicities. The two extra states correspond to a scalar (dilaton) and a pseudo-scalar (axion): the symmetric and antisymmetric combinations, respectively. This four-dimensional argument has a natural extension to any number of spacetime dimensions (with the axion substituted by the more general B-field), as we shall discuss. From these considerations, we see that the double copy of gauge theory will generically involve these additional fields, not just the graviton. The non-vacuum gravity solution under study in this paper, the Janis-Newman-Winicour solution, will provide a simple example of the inclusion of an additional field --- the dilaton.

Interactions in the `product-gravity' theory arise via the double copy from interactions in Yang-Mills theory, after stripping off the colour dependence of the latter in an appropriate manner. The rules of the double copy for scattering amplitudes provide a prescription for this. One question that has motivated much recent work is how to extend these ideas beyond scattering amplitudes, in particular to solutions of the classical equations of motion. The translation into this new setting is not trivial, and there are three obvious reasons for this. The first reason is that the solutions are typically expressed in coordinate space, rather than in momentum space, which is used for scattering amplitudes. The second reason is that scattering amplitudes exhibit gauge invariance, whereas explicit formulas for solutions depend on gauge choices. The third reason is that scattering amplitudes are studied in perturbation theory, and the rules of the double copy apply separately at each perturbative order. It is not clear a priori whether an exact solution in gravity can be expressed in a simple manner as a double copy of a gauge-theory solution.

Despite these difficulties, there is definite progress in relating exact solutions in gravity and gauge theory via the double copy. This is possible for a class of vacuum solutions in gravity, that of stationary Kerr-Schild spacetimes \cite{Monteiro:2014cda}, where much work has been done in this context; see e.g. \cite{Luna:2015paa,Ridgway:2015fdl,Luna:2016due,White:2016jzc,Bahjat-Abbas:2017htu,Carrillo-Gonzalez:2017iyj,DeSmet:2017rve,Berman:2018hwd,Gurses:2018ckx,Luna:2018dpt,Bahjat-Abbas:2018vgo,Andrzejewski:2019hub,CarrilloGonzalez:2019gof,Bah:2019sda,Alawadhi:2019urr,Banerjee:2019saj}. The basic example is the relation of the Schwarzschild and Coulomb solutions. More generally (in four dimensions), it applies to vacuum type-D spacetimes \cite{Luna:2018dpt}, relating, for instance, the C-metric to an analogous Lienard-Wiechert potential. These cases involve vacuum spacetimes, and yet we argued above that, more generally, the double copy should involve the dilaton and the B-field. An extension of the Kerr-Schild class of solutions that is well suited to deal with these fields was proposed in \cite{Lee:2018gxc}, based on the formalism of double field theory \cite{Siegel:1993xq,Siegel:1993th,Hull:2009mi,Hull:2009zb,Hohm:2010jy,Hohm:2010pp}. Several examples were considered with non-trivial configurations for the dilaton and the B-field. We will review and further extend this construction below. An extension to `heterotic gravity' was also given a double-copy interpretation \cite{Cho:2019ype}. These developments clearly demonstrate that the `double' in double copy and double field theory is indeed related; see \cite{Hohm:2011dz,Cheung:2016say} for earlier insights. Moreover, the double field theory approach explicitly relates the left-/right-moving factorisation associated to string theory, which is at the origin of the double copy, to the Kerr-Schild ansatz.

The progress in understanding solutions with dilaton and B-field raises a natural question. It has been argued that the double copy of a point charge (Coulomb) is not simply Schwarzschild, which is the example from the original Kerr-Schild double copy; it is an asymptotically-flat, static, spherically-symmetric solution containing a dilaton field \cite{Goldberger:2016iau,Luna:2016hge}. Ref.~\cite{Luna:2016hge} further argued that the double copy may be naturally defined to admit the dilaton or not, i.e., that it is not unique. The general double copy of a point charge was identified as the Janis-Newman-Winicour (JNW) solution \cite{Janis:1968zz}. The latter possesses two parameters, one associated to the graviton field and the other associated to the dilaton; the particular case of vanishing dilaton is the Schwarzschild solution. Here, we will present further arguments for why the double copy of a point charge is the full Janis-Newman-Winicour solution. Not only will we generalise the perturbative analysis of ref.~\cite{Luna:2016hge}, but we will actually present an exact map between the JNW solution and the Coulomb solution, along the lines of the Kerr-Schild double copy. This will involve extending the Kerr-Schild ansatz in double field theory, beyond the class of solutions considered in \cite{Lee:2018gxc}.

One apparent puzzle is that the `single-copy' gauge-theory solutions that are associated to the examples above are all Abelian. They trivialise the colour dependence. The solution to this puzzle is that many non-trivial gravity solutions (such as Schwarzschild or Kerr) are effectively linear, as is manifestly the case if they are of Kerr-Schild type. In fact, if we can write down an exact solution to the Einstein equations with a finite number of independent parameters, that solution should be `linear' in each parameter, in some sense. In this paper, we will see that this applies also to a solution, JNW, that deviates considerably from the Kerr-Schild property.

It is striking that a double-copy map between certain exact solutions, as explored here, is possible at all. Generically, the expected setting for the classical double copy is perturbative. This is how most discussions of the double copy for classical solutions have proceeded: the first approaches \cite{Saotome:2012vy,Neill:2013wsa,Monteiro:2011pc}, constructions based on the local symmetries \cite{Anastasiou:2014qba, Cardoso:2016ngt,Cardoso:2016amd,Luna:2016hge,Anastasiou:2018rdx,Borsten:2019prq}, use of the worldline formalism \cite{Goldberger:2016iau,Goldberger:2017frp,Goldberger:2017vcg,Chester:2017vcz,Goldberger:2017ogt,Li:2018qap,Carrillo-Gonzalez:2018pjk,Shen:2018ebu,Plefka:2018dpa,Plefka:2019hmz,Goldberger:2019xef}, and perturbation theory on curved backgrounds \cite{Adamo:2017nia,Adamo:2018mpq}. A double copy for classical observables (rather than solutions to the equations of motion) that follows more directly from that of scattering amplitudes has been explored with a view to gravitational phenomenology \cite{Bjerrum-Bohr:2014zsa,Bjerrum-Bohr:2016hpa,Luna:2017dtq,Kosower:2018adc,Chung:2018kqs,Bern:2019nnu,Bautista:2019tdr,Maybee:2019jus,Guevara:2019fsj,Arkani-Hamed:2019ymq,Johansson:2019dnu,Huang:2019cja,Bern:2019crd,Bautista:2019evw,Moynihan:2019bor,Kalin:2019rwq,Plefka:2019wyg}, a subject of obvious interest following the discovery of gravitational waves.

This paper is structured as follows. Section~\ref{sec:DCstory} contains a brief review and an overview of the paper. In section~\ref{sec:perturbative}, we use a perturbative construction to interpret the JNW solution as a double copy. In section~\ref{sec:exact}, we use an exact construction to interpret the JNW solution as a double copy, based on the formalism of double field theory. We conclude with a discussion of future directions in section~\ref{sec:conclusion}.


\section{Double-copy construction}
\label{sec:DCstory}

In the first part of this section, we present an overview of the double-copy construction, leaving greater detail for later sections. In the second part of this section, we review the JNW spacetime, whose double-copy relation to a point charge is the focus of our paper.

\subsection{Basics}

The starting point for the double copy is the tensor product structure between asymptotic states of perturbative scattering in Yang-Mills theory (stripped of colour) and gravity. The simplest example relates a product of polarisation vectors $\epsilon_\mu$ and $\tilde\epsilon_\nu$ to a polarisation tensor $\varepsilon_{\mu\nu}$, 
\begin{equation}
\label{eq:dcbasic}
\varepsilon_{\mu\nu} = \epsilon_\mu\, \tilde\epsilon_\nu\,.
\end{equation}
Since the polarisation vectors span a $(D-2)$-dimensional space, they induce a $(D-2)^2$-dimensional basis for the (generically non-factorisable) polarisation tensors in gravity. This gravity theory contains not only the graviton, but also an antisymmetric two-form field (B-field) and a scalar field (dilaton). The decomposition into these component fields can be written as
\begin{equation}
\label{eq:decvareps}
\varepsilon_{\mu\nu}=\varepsilon_{\mu\nu}^{(h)}+\varepsilon_{\mu\nu}^{(B)}+\varepsilon_{\mu\nu}^{(\phi)}\,,
\end{equation}
with
\begin{equation}
\varepsilon_{\mu\nu}^{(h)} = \varepsilon_{(\mu\nu)} - \frac{\Delta_{\mu\nu}}{d-2}\, \varepsilon_{\;\;\lambda}^\lambda\,, 
\qquad
\varepsilon_{\mu\nu}^{(B)} = \varepsilon_{[\mu\nu]}\,,
\qquad
\varepsilon_{\mu\nu}^{(\phi)} = \frac{\Delta_{\mu\nu}}{d-2}\, \varepsilon_{\;\;\lambda}^\lambda\,.
\end{equation}
The projector $\Delta_{\mu\nu}$ is associated to the completeness relation,
\begin{equation}
\label{completeness}
\sum^{D-2}_{r=1} \epsilon_\mu^{(r)} \,\epsilon_\nu^{(r)\ast} = \eta_{\mu\nu} - \frac{k_\mu q_\nu + k_\nu q_\mu}{k\cdot q} \equiv \Delta_{\mu\nu} \,.
\end{equation}
The low-energy interactions between the asymptotic states are basically fixed by gauge invariance, in both Yang-Mills theory and gravity. In the gravity case, the relevant theory (containing the dilaton and the B-field) is sometimes called NS-NS gravity, due to its appearance in the zero-mass level of the closed string, or alternatively $\mathcal N = 0$ supergravity. We shall discuss it further below. Remarkably, a factorisation reminiscent of \eqref{eq:dcbasic} exists for the full (interacting) theories, which follows from the relation between open and closed strings. In the context of scattering amplitudes $\mathcal A$, the double copy can be expressed in different formalisms, but we may schematically represent it as
\begin{equation}
\label{eq:dcSA}
{\mathcal A}_\text{\,NS-NS-grav} (\epsilon^{\,i}_{\mu} \tilde \epsilon^{\,i}_{\nu})\; =\;
{\mathcal A}_{\text{YM}} (\epsilon^{\,i}_{\mu}) \; { \otimes_\text{dc}}\; {\mathcal A}_{\text{YM}} (\tilde \epsilon^{\,i}_{\mu})\,.
\end{equation}
That is, the scattering amplitudes in gravity are given as a double copy of those in gauge theory. The `double-copy product' ${ \otimes_\text{dc}}$ involves inverse propagators and, of course, the stripping off of the colour dependence in the gauge theory amplitudes.

The remarks above have a coordinate-space analogue, where the double copy of Yang-Mills fields $A_\mu^a$ and $\tilde A_\mu^a$ leads to a gravity field $H_{\mu\nu}$, dubbed the `fat graviton', with $(D-2)^2$ degrees of freedom \cite{Luna:2016hge}. In this construction, to be reviewed in detail in section~\ref{subsec:linear}, the linearised `fat graviton' has a decomposition in terms of graviton, B-field and dilaton, analogous to \eqref{eq:decvareps}. In particular,
\begin{equation}
H_{\mu\nu} = \h_{\mu\nu} + B_{\mu\nu} + P^q_{\mu\nu} (\phi - \h)\,,
\end{equation}
where $P^q_{\mu\nu}$ is a coordinate-space version of $\Delta_{\mu\nu}/(D-2)$. To go beyond the free theory, one needs only the double copy. In particular, a Lagrangian for $H_{\mu\nu}$ can be constructed order by order in perturbation theory so as to obey the double copy for the scattering amplitudes \eqref{eq:dcSA}. The three-point interaction is explicitly discussed in section~\ref{subsec:nonlinear}, for instance. Starting from a  solution to the linearised theory, we can correct the solution order by order using those interactions. In sufficiently symmetric cases, one may be able to resum the perturbative solution, thereby obtaining an exact solution.

With this reasoning, the classical double copy is determined once we have the linearised classical double copy (at least for solutions that are continuously connected to the trivial solution). Consider the Coulomb solution in Yang-Mills theory,
\begin{equation}
\label{eq:coulomb}
A_{\mu}^a = -\frac{c^a}{r}\, u_\mu\,,
\end{equation}
where $u_\mu=(-1,0,0,0)$ and $\partial_\mu c^a=0$. The latter condition linearises the Yang-Mills equations, and therefore the solution is both linearised and exact. We may therefore substitute $c^a$ by a Maxwell electric charge. The question is now what is its double copy. A natural guess is\footnote{We will define later our normalisation conventions.}
\begin{equation}
H_{\mu\nu} = \frac{\kappa\,M}{8\pi\,r}\,u_\mu u_\nu \,,
\end{equation}
which solves the linearised equations of motion. Notice that the B-field vanishes as $H_{\mu\nu}$ is symmetric. Given that this is a static, spherically symmetric and asymptotically flat solution to the Einstein equations with a minimally-coupled scalar field (the dilaton), the exact solution is unique and known explicitly. It is the JNW solution, to be reviewed in the next subsection. This solution possesses two parameters, $M$ and $Y$, which are associated respectively to the graviton field and the dilaton. The case above corresponds to the solution with $M=Y$. The double-copy interpretation of this solution was also discussed in \cite{Goldberger:2016iau}, based on a worldline formalism for constructing solutions. For reference, the `fat graviton' field (at linearised level) for the JNW solution can be written as \begin{equation}
\label{eq:HJNW}
H_{\mu\nu} =  \frac{\kappa}{2} \frac{1}{4\pi r}\left(M \,u_\mu u_\nu + (M-Y) \;\frac{1}{2} (\eta_{\mu\nu} - q_\mu l_\nu - q_\nu l_\mu) \right) \,,
\end{equation}
where $q_\mu=(1,0,0,1)$ and $l_\mu=(0,x,y,r+z)/(r+z)$ \cite{Luna:2016hge}.

There is another natural guess for the double copy of the point charge, though --- the Schwarzschild solution. Suppose that we write the Coulomb solution in a different gauge,
\begin{equation}
\label{eq:coulombKS}
{A'}_{\mu}^a = -\frac{c^a}{r}\, k_\mu\,,
\end{equation}
where $k_\mu=(1,\mathbf{x}/r)$. Then a natural guess for the corresponding metric is
\begin{equation}
\label{eq:KSmetric}
g_{\mu\nu} = \eta_{\mu\nu} + \frac{\kappa^2 M}{8\pi\,r}\, k_\mu k_\nu\,.
\end{equation}
This turns out to be the Schwarzschild solution in Kerr-Schild coordinates. So now we have a vacuum solution, i.e., no dilaton. Indeed, the Schwarzschild solution is a particular case of the JNW solution, the one where $Y=0$. The metric exhibits in Kerr-Schild coordinates the property that it is both linearised and exact. This property was instrumental in interpreting the exact solution as the double copy of a point charge in \cite{Monteiro:2014cda}, and this conclusion extends to many other cases, including the Kerr and Taub-NUT metrics --- in fact, it extends to all vacuum type D spacetimes \cite{Luna:2018dpt}. This double-copy interpretation is further supported by more recent arguments involving solution-generating techniques, computations with scattering amplitudes, duality considerations, and asymptotic symmetries \cite{Arkani-Hamed:2019ymq,Godazgar:2019ikr,Moynihan:2019bor,Huang:2019cja,Alawadhi:2019urr}. Notice that, in terms of the `fat graviton' \eqref{eq:HJNW}, the Schwarzschild case ($Y=0$) does not look particularly simple.

Each of the examples looks more natural for certain coordinates or field choices. They are also consistent with the linearised double-copy ideas developed in \cite{Anastasiou:2014qba,Cardoso:2016amd,Cardoso:2016ngt,Anastasiou:2018rdx}.\footnote{This is based on a convolution with a certain scalar field, related to the bi-adjoint scalar. With spherical symmetry, the procedure roughly amounts to $1/r=(1/r)\ast \text{inv}(1/r) \ast (1/r)$, where, on the right-hand side, the first and third factors come from the pair of Coulomb solutions, and in the middle factor ``$\text{inv}$" denotes an inverse with respect to the convolution $\ast$. This justifies the fact that the exact classical double copy works locally in coordinate space for a special class of solutions (and gauges), whereas in general one expects the double copy to work locally only in momentum space.} We argue, following \cite{Luna:2016hge}, that the general JNW solution should be interpreted as the double copy of a point charge. In fact, we will find in this paper an exact double-copy map from the JNW solution to the Coulomb solution, along the lines of the Kerr-Schild double copy (even though the solution is not of Kerr-Schild type). Of the two parameters of the JNW solution, $M$ and $Y$, only a combination survives in the `single copy', associated to the charge parameter of the Coulomb solution. Logically, the reverse path --- from Coulomb to JNW --- must allow for the introduction of an additional parameter, which distinguishes $M$ and $Y$, as argued above.

From our initial considerations with polarisation vectors and tensors, the analogue statement is that, with a pair of polarisation tensors $\epsilon_\mu$ and $\tilde\epsilon_\mu$, it is natural to consider different tensorial structures for the `product', namely $\epsilon_{(\mu}\tilde\epsilon_{\nu)}$, $\epsilon_{[\mu}\tilde\epsilon_{\nu]}$, and also $\Delta_{\mu\nu}\,\epsilon\cdot\tilde\epsilon$. With a single polarisation vector in hand, say $\epsilon_\mu$, the most general double copy is the combination
\begin{equation}
C^{(h)} \left(\epsilon_\mu \epsilon_\nu - \frac{\Delta_{\mu\nu}}{d-2}\,\epsilon\cdot\epsilon\right) + C^{(\phi)}  \,\frac{\Delta_{\mu\nu}}{d-2}\,\epsilon\cdot\epsilon\,,
\end{equation}
where $C^{(h)}$ and $C^{(\phi)}$ are the two parameters. This is the analogue of the JNW solution. With two distinct polarisation tensors, $\epsilon_\mu$ and $\tilde\epsilon_\mu$, it is also natural to tune the B-field component.

\subsection{The JNW solution}
\label{subsec:JNW}

We are interested in a static, spherically symmetric solution first obtained by Janis, Newman and Winicour (JNW)~\cite{Janis:1968zz}. This solution to the Einstein equations with a minimally-coupled scalar reads
\begin{align}
\df s^2 & = - \left(1 - \frac{\rho_0}{\rho}\right)^\gamma \df t^2 +\left(1 - \frac{\rho_0}{\rho}\right)^{-\gamma} \df \rho^2 + \left(1 - \frac{\rho_0}{\rho}\right)^{1-\gamma} \rho^2 \df\Omega^2,
\label{JNWds2} \\
\phi & = \frac{\kappa}{2} \frac{Y}{4\pi \rho_0} \log \left( 1 - \frac{\rho_0}{\rho} \right).
\label{JNWphi}
\end{align}
The two parameters $\rho_0$ and $\gamma$ can be given in terms of the mass $M$ and the scalar coupling $Y$ as
\begin{equation}
\rho_0 = 2 G \sqrt{M^2 + Y^2}= \left(\frac{\kappa}{2}\right)^2 \frac{\sqrt{M^2 + Y^2}}{4\pi}, \qquad
\gamma = \frac{M}{\sqrt{M^2 + Y^2}}.
\label{JNWMY}
\end{equation}
The special case for which $Y=0$ and $M>0$ (and therefore $\gamma=1$) is the Schwarzschild solution. If $M>0$ but the dilaton field is non-vanishing, i.e., $|Y|>0$, then the solution is still asymptotically flat, but there is a naked singularity at zero radius, which corresponds to $\rho=\rho_0$ since the 2-sphere factor vanishes in the line element. This naked singularity is not surprising because the uniqueness theorems prevent a scalar-hair deformation of the Schwarzschild solution.

The JNW solution is a natural point charge solution in the double-copy gravity theory, i.e., NS-NS gravity, so it is not surprising that it is related to the Coulomb solution. In the exact double-copy construction of section~\ref{sec:exact}, we will find that, for any solution to NS-NS gravity that respects a certain Kerr-Schild-inspired ansatz, there are two corresponding solutions to the Maxwell equations,
\begin{equation}
\mathcal{R}_{\mu\nu}=0 \qquad \Rightarrow \qquad \partial^\mu F_{\mu\nu}=0\,, \quad \partial^\mu \bar F_{\mu\nu}=0\,.
\label{}\end{equation}
Here, $\mathcal{R}_{\mu\nu}$ is a double field theory analogue of $R_{\mu\nu}$ for vacuum gravity. The Maxwell solutions with field strength $F_{\mu\nu}$ and $\bar F_{\mu\nu}$ are the pair whose double copy is the NS-NS gravity solution, and they are associated respectively to left and right movers from a string theory interpretation. For the JNW solution, $F_{\mu\nu}=\bar F_{\mu\nu}$, and they correspond to Coulomb. In the vacuum case, where JNW reduces to Schwarzschild, this coincides with the original Kerr-Schild double copy.


\section{Perturbative double copy}
\label{sec:perturbative}

In this section, we follow a perturbative construction of the classical double copy presented in \cite{Luna:2016hge}. In that work, this construction was illustrated by a particular case of the JNW family, with $M=Y$, for which the calculations are easier. The aim of this section is to apply the construction to a generic JNW solution, with $M\neq Y$. Below, we review the formalism and its application to the linearised JNW solution, before studying the next order in perturbation theory.

\subsection{Linear level: review}
\label{subsec:linear}

The basic object in the construction of Ref.~\cite{Luna:2016hge} is the gravity field that naturally arises as the double copy of Yang-Mills theory, denoted by $H_{\mu\nu}$, which was dubbed the `fat graviton'. This field is a massless tensor with $(D-2)^2$ degrees of freedom. It provides an alternative formulation, at least in perturbation theory, to Einstein gravity coupled to a dilaton field $\phi$ and a two-form field $B_{\mu\nu}$. The latter is known as B-field or Kalb-Ramond field in the context of string theory, where this gravity theory arises in the low energy limit of the closed bosonic string. The action is
\begin{equation}
S=\int d^D x\sqrt{-g}\left[\frac{2}{\kappa^2}R
-\frac{1}{2(D-2)} (\partial \phi)^2
-\frac{1}{6}e^{-2\kappa\phi/{(D-2)}}(\df B)^2\right]\,,
\label{EinsteinAction}
\end{equation}
where $(dB)_{\mu\nu\lambda}$ is the field strength for $B_{\mu\nu}$. We use a particular normalisation of the fields that simplifies some of the constant coefficients that appear in the perturbation theory. The field associated to the metric may be expressed as the `gothic graviton' $\h_{\mu\nu}$ via\footnote{At linearised level, this coincides with the `trace-reversed graviton', \,$
\h_{\mu\nu} = h_{\mu\nu} - \frac12 \eta_{\mu\nu} h$\,, where $h_{\mu\nu} $ is the usual metric perturbation and $h$ is its trace.}
\begin{equation}
\sqrt{-g}\,g^{\mu\nu} = \eta^{\mu\nu} - \kappa\, \h^{\mu\nu}\,,
\label{gothich}
\end{equation}
for which the de Donder gauge is simply given by \,$\partial_\mu \h^{\mu\nu} = 0$\,.
Following \cite{Luna:2016hge}, we refer to $\h_{\mu\nu}$, $\phi$ and $B_{\mu\nu}$ as the `skinny fields', in contrast to the representation of the full content of these fields in terms of the `fat graviton' $H_{\mu\nu}$. Together, the `skinny fields' have $(D-2)^2$ perturbative degrees of freedom, the same as the `fat graviton'.

Let us recall how this dictionary works at linearised level. Consider the de Donder gauge for the graviton field, and a Lorentz-type gauge for the B-field, $\partial^\mu B_{\mu\nu}=0$. Then the linearised equations of motion for the `skinny fields' are simply
\begin{equation}
\partial^2 \h_{\mu\nu}=0\,, \qquad 
\partial^2 \phi=0\,, \qquad 
\partial^2 B_{\mu\nu}=0\,.
\label{EOMskinny}
\end{equation}
The `fat graviton' is defined, at linearised level, as
\begin{equation}
H_{\mu\nu} = \h_{\mu\nu} + B_{\mu\nu} + P^q_{\mu\nu} (\phi - \h)\,,
\label{eq:linearFatDef}
\end{equation}
where $\h=\eta^{\mu\nu}\h_{\mu\nu}$ and $P^q_{\mu\nu}$ is a coordinate-space realisation of a momentum-space projector, which we will define momentarily. The `fat graviton' is defined in this way so that it satisfies both the Lorentz-type condition \,$\partial^\mu H_{\mu\nu} = 0$ and the equation of motion
\begin{equation}
\partial^2 H_{\mu\nu}=0\,.
\label{EOMfat}
\end{equation}
Therefore, this field has a simple propagator. In order to incorporate the rules of the perturbative (BCJ) double copy as they appear in scattering amplitudes \cite{Bern:2008qj}, the next step is to write the interaction vertices appropriately, in terms of those of a BCJ-Lagrangian for Yang-Mills theory \cite{Bern:2010yg,Tolotti:2013caa}. We will see the simplest example in the next subsection, for the three-point vertex.

The relation \eqref{eq:linearFatDef} can be easily inverted, so that
\begin{align}
\phi &= H^\mu{}_\mu \equiv H, \\
B_{\mu\nu} &= \frac12 \left( H_{\mu\nu} - H_{\nu\mu} \right), \\
\h_{\mu\nu} - P^q_{\mu\nu} \h &= \frac12 \left( H_{\mu\nu} + H_{\nu\mu} \right) - P^q_{\mu\nu} H\,.
\end{align}
It turns out that $\h_{\mu\nu}$ and $\h_{\mu\nu} - P^q_{\mu\nu} \h$ differ only by a diffeomorphism that does not affect $\phi$ and $B_{\mu\nu}$ to this order in perturbation theory; see \cite{Luna:2016hge} for more details.

The crucial object in the construction above is $P^q_{\mu\nu}$, which is defined as a non-local operator,
\begin{equation}
P^q_{\mu\nu} = \frac{1}{D-2}\left( \eta_{\mu\nu} - \frac{q_\mu \partial_\nu + q_\nu \partial_\mu}{q \cdot \partial} \right)\,.
\label{projectorq}
\end{equation}
Here, $q_\mu$ is a null reference vector which we take to be constant to avoid ambiguities. The role of $P^q_{\mu\nu}$ in \eqref{eq:linearFatDef} is best understood by taking the `fat graviton' to be a linearised plane wave with polarisation tensor $\varepsilon_{\mu\nu}$, obeying the transversality condition $\,k^\mu\varepsilon_{\mu\nu}=k^\mu\varepsilon_{\nu\mu}=0$\,, and the gauge fixing condition \,$ q^\mu\varepsilon_{\mu\nu}=q^\mu\varepsilon_{\nu\mu}=0$\,. Then, the definition \eqref{eq:linearFatDef} is simply a coordinate-space version of the decomposition of this polarisation tensor into the `skinny field' polarisations, discussed previously in \eqref{eq:decvareps}. The properties of the projector, $\,\Delta_\lambda^{\lambda}=D-2\,$ and $\,\Delta^{\lambda}_\mu\, \Delta^{\nu}_\lambda=\Delta^{\nu}_\mu$\,, are inherited by $(D-2)P^q_{\mu\nu}$. We use the symbol $q$ in $P^q_{\mu\nu}$ to keep in mind the gauge dependence.


The application of this story to the four-dimensional JNW solution is straightforward. First, we extract from the solution \eqref{JNWds2}-\eqref{JNWphi} the `skinny fields'. Moving to de Donder gauge via the coordinate transformation \,$\rho=r+\rho_0/2$\,, we have
\begin{equation}
\h^{(0)}_{\mu\nu} = \frac{\kappa}{2}\frac{M}{4\pi\,r}\,u_\mu u_\nu \,, \qquad \phi^{(0)}=-\frac{\kappa}{2}\frac{Y}{4\pi\,r} \,, \qquad B^{(0)}_{\mu\nu} = 0\,,
\end{equation}
where $u^\mu=(1,0,0,0)$\,. We have now introduced a superscript ${}^{(0)}$ to indicate that these are the leading order contributions (in $\kappa$), i.e., the linearised fields, as opposed to the next-to-leading order correction considered in the next subsection. The linearised fat graviton then reads
\begin{equation}
\begin{split}
H^{(0)}_{\mu\nu}=&\, \h^{(0)}_{\mu\nu}-P^q_{\mu\nu}\h^{(0)}+P^q_{\mu\nu}\phi^{(0)}
\\=&\, \frac{\kappa}{2}\frac{1}{4\pi\,r}\left(M \,u_\mu u_\nu + (M-Y)P_q^{\mu\nu}\left(\frac{1}{r}\right)\right)
\\=& \,\frac{\kappa}{2} \frac{1}{4\pi r}\left(M \,u_\mu u_\nu + (M-Y) \;\frac{1}{2} (\eta_{\mu\nu} - q_\mu l_\nu - q_\nu l_\mu) \right).
\end{split}
\end{equation} 
The last step is the result of a computation in which the null reference vector was chosen to be $q_\mu=(1,0,0,1)$, yielding $l_\mu=(0,x,y,r+z)/(r+z)$, with $q\cdot l=1$.

\subsection{Beyond linear level}
\label{subsec:nonlinear}

Now we will discuss the first correction to the linearised theory. We start by presenting the formalism, and then consider the JNW case. For simplicity, we set the B-field to vanish, so that the fat graviton is symmetric.

\subsubsection{Next-to-linear order formalism}
In order to get the next-to-linear order term for the fat graviton, we apply the double copy procedure: the three-point interaction is the double copy of the kinematic part of the Yang-Mills three-point interaction. That is the whole point of working with the fat graviton \cite{Luna:2016hge}. In momentum space,
\begin{align}
H^{(1)\mu\mu^\prime}(-p_1)=\frac{1}{4\, p_1^2}\int &\dbar^D p_2\,\dbar^D\, p_3 \deltabar^D(p_1+p_2+p_3)\nonumber\\
&\times \left[ (p_1-p_2)^{\gamma}\eta^{\mu\beta}+(p_2-p_3)^{\mu}\eta^{\beta\gamma}+(p_3-p_1)^{\beta}\eta^{\mu\gamma}  \right] \\
&\times \left[ (p_1-p_2)^{\gamma^\prime}\eta^{\mu^\prime\beta^\prime}+(p_2-p_3)^{\mu^\prime}\eta^{\beta^\prime\gamma^\prime}+(p_3-p_1)^{\beta^\prime}\eta^{\mu^\prime\gamma^\prime}  \right]H^{(0)}_{\beta\beta^\prime}(p_2)H^{(0)}_{\gamma\gamma^\prime}(p_3)\nonumber~.
\end{align}
Notice how the unprimed and primed indices only contract with each other. The shorthand notation for differentials and deltas is
\begin{equation}
\deltabar^D(x)\equiv(2\pi)^D\delta^{(D)}(x)~,\qquad \int\dbar^D p\,F(p)\equiv\int\frac{d^Dp}{(2\pi)^D}F(p)~.
\end{equation} 
Note that, due to the integration, we have a symmetry $2\leftrightarrow 3$.
Introducing the notation $H^{(0)}_{\mu\nu}(p_i)=H^{(0)}_{i\,\mu\nu}$ and using dots to denote the contractions,\footnote{The dots indicate the contractions in the sense: $p\cdot H^{\mu}=p_{\nu}\,H^{\nu\mu}$, 
$ \quad H^{\mu}\cdot p=\,H^{\mu\nu}p_{\nu}$, 
$\quad p\cdot H_i\cdot H_j^{\mu}=p_{\sigma}\,H^{\sigma}_{j\,\nu}H^{\nu\mu}_j$ and 
$H_i\cdot H_j=H_i^{\mu\nu}H_{j\, \nu\mu}$. }
\begin{equation}
\begin{split}
H^{(1)\mu\mu^\prime}(-p_1)&=\frac{1}{4\, p_1^2}\int \dbar^D p_2\,\dbar^D\, p_3 \deltabar^D(p_1+p_2+p_3)\bigg{[}\\
&H^{(0)\mu\mu^\prime}_2(p_1-p_2)\cdot H^{(0)}_3\cdot(p_1-p_2)
-H_2^{(0)\mu}\cdot(p_1-p_3)\,(p_1-p_2)\cdot H_3^{(0)\mu^\prime}\\
&+(p_2-p_3)^{\mu}(p_1-p_2)\cdot H^{(0)T}_3\cdot H^{(0)\mu^\prime}_2
+ (p_2-p_3)^{\mu^\prime}(p_1-p_2)\cdot H^{(0)}_3\cdot H^{(0)T\mu}_2\, \\
&+\frac{1}{2}(p_2-p_3)^\mu(p_2-p_3)^{\mu^\prime} \,H^{(0)}_2\cdot H^{(0)T}_3
+(2\leftrightarrow3)\bigg{]}~.
\end{split}
\end{equation}
Let us now set the B-field to zero, i.e., take $H^{(0)}_{\mu\nu}$ to be symmetric. Using the linear condition of transversality, \, $p_i\cdot H^{(0)}_i=0$\,, and momentum conservation, we obtain
\begin{align}
H^{(1)\mu\mu^\prime}(-p_1)&=\frac{1}{4\, p_1^2}\int \dbar^D p_2\,\dbar^D\, p_3 \deltabar^D(p_1+p_2+p_3)\bigg{[}\nonumber \\
&4\,H^{(0)\mu\mu^\prime}_2 p_2\cdot H^{(0)}_3\cdot p_2
-4\,p_2^{(\mu}H^{(0)\mu^\prime)}_2\cdot H^{(0)}_3\cdot p_2
+4\,p_3^{(\mu}H^{(0)\mu^\prime)}_2\cdot H^{(0)}_3\cdot p_2 \label{eq:H1 from H0}\\
&-4\,H_2^{(0)(\mu}\cdot p_3\,H_3^{(0)\mu^\prime)}\cdot p_2
+\frac{1}{2}(p_2-p_3)^\mu(p_2-p_3)^{\mu^\prime} \,H^{(0)}_2\cdot H^{(0)}_3+(2\leftrightarrow3)\bigg{]}~.\nonumber
\end{align}
Alternatively, it can be written directly in terms of the linearised skinny fields,
\begin{equation}
\begin{split}
H^{(1)\mu\mu^\prime}(-p_1)&=\frac{1}{4\, p_1^2}\int \dbar^D p_2\,\dbar^D\, p_3 \deltabar^D(p_1+p_2+p_3)\bigg{[} \\
&4\,\h^{(0)\mu\mu^\prime}_2 p_2\cdot \h^{(0)}_3\cdot p_2
-4\,p_2^{(\mu}\h^{(0)\mu^\prime)}_2\cdot \h^{(0)}_3\cdot p_2
+4\,p_3^{(\mu}\h^{(0)\mu^\prime)}_2\cdot \h^{(0)}_3\cdot p_2 \\
&-4\,\h_2^{(0)(\mu}\cdot p_3\,\h_3^{(0)\mu^\prime)}\cdot p_2
+\frac{1}{2}(p_2-p_3)^\mu(p_2-p_3)^{\mu^\prime} \,\h_2\cdot \h_3
+X^{\mu\mu^\prime}
+(2\leftrightarrow3)\bigg{]}~,
%
\end{split}\label{eq:H1 from h0, before T}
\end{equation}
where $X^{\mu\mu'}
$ accounts for all the terms involving the projectors. So we have the general expression for the fat graviton at this order, either in terms of the linearised fat graviton, or in terms of the linearised skinny fields.

Suppose now that want to compare this directly with the JNW solution at next-to-leading order. This is, of course, against the spirit of the double copy. In accordance with this spirit, we would solve a complete problem in terms of the fat graviton description only, for which we have the simple propagator and the interaction rules determined by the Yang-Mills rules. However, it may be useful for certain purposes to provide the dictionary between the fat graviton and the skinny fields also at next-to-leading order. This is cumbersome, because we have to keep track of gauge choices and field redefinitions, which do not follow trivially from the linear case. As in \cite{Luna:2016hge}, we can build a {transformation tensor} $\mathcal{T}_{\mu\nu}$ that represents all the gauge transformations and field redefinitions needed to retrieve the skinny fields in their standard form -- in particular, so that the gothic graviton is in de Donder gauge. The dictionary at this order is
\begin{equation}
\begin{split}
H^{(1)\mu\nu}=&\h^{(1)\mu\nu}-P_{q^\prime}^{\mu\nu}(\h^{(1)}-\phi^{(1)}) +\mathcal{T}^{(1)\mu\nu}~,\label{eq:H1 from h1}
\end{split}
\end{equation} 
where the transformation function can be written, for instance, in terms of the linearised skinny fields,\footnote{This expression is not exactly the one presented in \cite{Luna:2016hge} because that was simplified to be valid only in the special JNW case where $M=Y$.}
\begin{equation}
\begin{split}
\mathcal{T}^{(1)\mu\mu^\prime}(-p_1)&=\frac{1}{4\, p_1^2}\int \dbar^D p_2\,\dbar^D\, p_3 \deltabar^D(p_1+p_2+p_3)\bigg{[} \\
&\h_{2}^{(0)}\cdot \h_{3}^{(0)}\left(\frac{1}{2}p_1^\mu\,p_1^{\mu^\prime}-\eta^{\mu\mu^\prime}p_2\cdot p_3\right)
+4\,p_1^{(\mu}\h^{(0)\mu^\prime)}_2\cdot\h_3^{(0)}\cdot p_2\\
&+4\,\h_2^{(0)(\mu}\cdot\h_3^{(0)\mu^\prime)}p_2\cdot p_3
+2\eta^{\mu\mu^\prime}p_3\cdot \h^{(0)}_2\cdot\h^{(0)}_3\cdot p_2\\
&+\frac{1}{2}\left(
\eta^{\mu\mu^\prime}p_2\cdot p_3- 2p_2^\mu p_3^{\mu^\prime}\right)\left(-\h^{(0)}_{2} \h^{(0)}_{3}+\phi^{(0)}_2\phi^{(0)}_3\right)
+X^{\mu\mu^\prime}\\
&+P_{q^\prime}^{\mu\nu}\bigg{(}(D-6)p_2\cdot p_3\, \h_{2}^{(0)}\cdot \h_{3}^{(0)}
-2(D-2)p_3\cdot\h^{(0)}_2\cdot \h^{(0)}_3\cdot p_2
\\
&-4\,(\phi^{(0)}_2-\h^{(0)}_2)\,p_2\cdot\h_3^{(0)}\cdot p_2-\frac{D-2}{2}p_2\cdot p_3\left(-\h^{(0)}_{2} \h^{(0)}_{3}+\phi^{(0)}_2\phi^{(0)}_3\right)+(2\leftrightarrow3)\bigg{)}\bigg{]}~.
\end{split}\label{eq:T1 from H0}
\end{equation}
We allow the reference null vector $q'$ here to be different from the linear-level discussion since this clarifies the extraction of the skinny fields from the fat graviton. This proceeds as follows.
From \eqref{eq:H1 from h1}, it can be checked that 
\begin{equation}
\h^{(1)\mu\nu}-P_{q^\prime}^{\mu\nu}(\h^{(1)})=H^{(1)\mu\nu}-\mathcal{T}^{(1)\mu\nu}-P_{q^\prime}^{\mu\nu}(H^{(1)}-\mathcal{T}^{(1)})\,.
\end{equation}
Substituting \eqref{eq:H1 from H0} and \eqref{eq:T1 from H0}, and comparing the terms without dependence on the auxiliary vector $q^\prime$, we get the general result for the graviton in the desired gauge
\begin{equation}
\begin{split}
\h^{(1)\mu\nu}&=-\frac{1}{4\, p_1^2}\int \dbar^D p_2\,\dbar^D\, p_3 \deltabar^D(p_1+p_2+p_3)\bigg{[} \\
& 4 p_2\cdot p_3 \h^{(0)(\mu}_{2} \cdot \h^{(0)\nu)}_{3}
 +2  \h^{(0)}_{2}\cdot \h^{(0)}_{3}p_2^{(\mu} p_3^{\nu)} 
 -4\h^{(0)\mu\nu}_{3 } p_3\cdot \h^{(0)}_{2 } \cdot p_3  
\\&
+4p_2\cdot \h^{(0)(\mu}_{3} p_3\cdot \h^{(0)\nu)}_{2}
-8p_2\cdot \h^{(0)}_{3}\cdot \h^{(0)(\mu}_{2 }p_3^{\nu)}
\\& 
-\eta^{\mu\nu}\h^{(0)}_{2}\cdot \h^{(0)}_{3}p_2\cdot p_3
+2\eta^{\mu\nu}p_2\cdot \h^{(0)}_{3}\cdot \h^{(0)}_{2}\cdot p_3
\\
&-\frac{1}{2}\left(
\eta^{\mu\nu}p_2\cdot p_3 - 2p_2^{(\mu} p_3^{\nu)}\right)\left(-\h^{(0)}_{2} \h^{(0)}_{3}+\phi^{(0)}_2\phi^{(0)}_3\right)+(2\leftrightarrow 3)\bigg{]}~.
\end{split}
\label{gothgrav1}
\end{equation}
The dilaton is obtained in a similar way,
\begin{equation}
\begin{split}
\phi^{(1)}&=H^{(1)}-\mathcal{T}^{(1)}=\\
&\frac{1}{4\, p_1^2}\int \dbar^D p_2\,\dbar^D\, p_3 \deltabar^D(p_1+p_2+p_3)\bigg{[}4\phi_2^{(0)}\,p_2\cdot\h^{(0)}_3\cdot p_2+(2\leftrightarrow 3)\bigg{]} ~.
\end{split}\label{eq:Phi1 from h0}
\end{equation}

\subsubsection{JNW case}

For illustration, let us apply the general construction to the four-dimensional JNW case. In de Donder gauge, and to the relevant order, the JNW solution is given by 
\begin{equation}
\begin{split}
\h_{\mu\nu}&=\frac{\kappa}{2}\frac{M}{4\pi\,r}u_\mu u_\nu
+\left(\frac{\kappa}{2}\right)^3 \frac{1}{8(4\pi)^2\,r^2}
\left(
(7M^2-Y^2){u_\mu u_\nu}+(M^2+Y^2)\,\hat{r}_\mu \hat{r}_\nu
\right)+\mathcal{O}(\kappa^5)\\
\phi&=-\frac{\kappa}{2}\frac{Y}{4\pi\,r}+\mathcal{O}(\kappa^5)~,
\end{split}\label{eq:JNW de Donder}
\end{equation}
where $\hat{r}^\mu=(0,\mathbf{x}/r)$ and $u^\mu=(1,0,0,0)$. We can reproduce these expressions with the formalism explained above. The starting point is the linear graviton and dilaton, which are taken to be
\begin{equation}
\begin{split}
\h^{(0)\mu\nu}=&\,\frac{\kappa}{2} M\frac{ u^\mu u^\mu}{4\pi r}\quad\longrightarrow\quad \frac{\kappa}{2} M\frac{ u^\mu u^\mu \deltabar^1(p^0)}{p^2}  ~,
\\
\phi^{(0)}=&-\frac{\kappa}{2}Y\frac{1}{4\pi r}   \quad\longrightarrow\quad -\frac{\kappa}{2}Y\frac{\deltabar^1(p^0)}{p^2}~,
\end{split}
\end{equation} 
in coordinate space and momentum space, respectively.
The expression \eqref{gothgrav1} for the graviton is simplified by the fact that $p_i\cdot u=0$, and we have
\begin{equation}
\begin{split}
\h^{(1)\mu\nu}
&=\frac{-1}{4\,p_1^2}\left(\frac{\kappa}{2}\right)^3\int \dbar^4p_2\dbar^4p_3 \deltabar^{(4)}(p_1+p_2+p_3)\frac{\deltabar(p_2^0)}{p_2^2}\frac{\deltabar(p_3^0)}{p_3^2}
\bigg\{
-8\,M^2\,p_2\cdot p_3 u^\mu u^\nu\\
&+4\,M^2p_2^{(\mu}p_3^{\nu)}
-2\,M^2p_2\cdot p_3\,\eta^{\mu\nu}
-\left(\eta^{\mu\nu}p_2\cdot p_3-2p_2^{(\mu}p_3^{\nu)}\right)(Y^2-M^2)
\bigg\}
\,.
\end{split}
\end{equation}
In order to invert back to position space, we use the identities
\begin{equation}
\begin{split}
\int \dbar^4p_2\dbar^4p_3 \deltabar^{(4)}(p_1+p_2+p_3)\frac{\deltabar^1(p_2^0)\deltabar^1(p_3^0)}{p_2^2p_3^2} \frac{p_2^{\mu}p_3^{\nu}}{p_1^2}& = \frac{-1}{4(4\pi)^2}\left[\frac{x^\mu x^\nu}{r^4}-\frac{\eta_{(3)}^{\mu\nu}}{r^2}\right]~,\\
\Rightarrow\quad \int \dbar^4p_2\dbar^4p_3 \deltabar^{(4)}(p_1+p_2+p_3)\frac{\deltabar^1(p_2^0)\deltabar^1(p_3^0)}{p_2^2p_3^2} \frac{p_2\cdot p_3}{p_1^2}&=\frac{1}{2(4\pi)^2}\frac{\eta_{(3)}^{\mu\nu}}{r^2}~,
\end{split}
\end{equation}
where we have defined $\eta_{(3)}^{\mu\nu}=\eta^{\mu\nu}+u^\mu u^\nu$.\footnote{These identities can be derived in a similar way to (56) in \cite{Luna:2016hge}.} We obtain
\begin{equation}
\begin{split}
\h^{(1)\mu\nu}=&\,\frac{-1}{16(4\pi)^2}\left(\frac{\kappa}{2}\right)^3
\bigg\{
\left(-8\,M^2 u^\mu u^\nu -(M^2+Y^2)\eta^{\mu\nu}\right)\frac{2}{r^2}+2(M^2+Y^2)\left(\frac{\eta_{(3)}^{\mu\nu}}{r^2}-\frac{x^\mu x^\nu}{r^4}\right)
\bigg\}
\\=&\,\frac{1}{8(4\pi)^2}
\left(\frac{\kappa}{2}\right)^3\bigg\{ (7M^2-Y^2)\frac{u^\mu u^\nu}{r^2} +(M^2+Y^2)\frac{x^\mu x^\nu}{r^4}
\bigg\}~.
\end{split}
\end{equation}
This result matches \eqref{eq:JNW de Donder}. As for the dilaton, the correction \eqref{eq:Phi1 from h0}  vanishes by virtue of $p_i\cdot u=0\Rightarrow p_i\cdot \h^{(0)}_j=0$, which also matches \eqref{eq:JNW de Donder}.


\section{Exact double copy}
\label{sec:exact}

So far we have considered the perturbative double copy for the JNW solution. We now discuss how to define an exact double-copy map based on double field theory (DFT). To this end, we introduce an ansatz for generalised metric in DFT, by relaxing the null condition in the Kerr-Schild (KS) formalism, and derive a pair of Maxwell solutions as the two factors in the double copy. We apply this general formalism to the JNW case and show that both Maxwell solutions are the Coulomb potential, which is therefore the `single copy' of JNW.

\subsection{Rewriting the JNW solution}

We have introduced the JNW metric in a previous section. That metric solves the Einstein-dilaton equations of motion in the Einstein frame. However, for the remaining sections it will be more convenient to work in the string frame. This is achieved by performing the following field redefinition
\begin{equation}
\begin{gathered}
g^E_{\mu\nu}\ \rightarrow\  g^S_{\mu\nu}
=e^{\sigma (\phi-\phi_0)} g^E_{\mu\nu}~,\\
\phi_0=\lim_{r\rightarrow \infty}\phi~,\\
\end{gathered}
\end{equation}
where the constant $\sigma$ depends on the choice of normalisation for the dilaton. In this section, since we are not working in perturbation theory, we will suppress the coupling constant $\kappa$, and use instead a common string-frame normalisation convention for the fields.
The action in the string frame then reads
\begin{equation}
S=\int d^4x \sqrt{-g^s}e^{-2\phi}\left(R-\frac{1}{12}H_{\mu\nu\rho}H^{\mu\nu\rho}+4\,\partial_\mu\phi\,\partial^\mu\phi\right)~,
\label{eq:stringframeaction}
\end{equation}
which corresponds to the low energy effective action of string theory. In the string frame, the JNW metric is given by
\begin{equation}
\begin{gathered}
\df s^2=e^{2\phi}\left[-\left(1-\frac{r_0}{r}\right)^{\frac{a}{r_0}}\df t^2
+\left(1-\frac{r_0}{r}\right)^{\frac{-a}{r_0}}
\left(\df r^2+r(r-r_0)\,\df \Omega_2^2\,\right)\right]~, \\
e^{2\phi}=\left(1-\frac{r_0}{r}\right)^\frac{b}{r_0}~, \qquad
r_0=\sqrt{a^2+b^2}~, 
\end{gathered}\label{eq:JNW in sf}
\end{equation}
and $\phi_0=0$. Apart from different normalisation conventions, $a$ is $M$, and $b$ is $Y$, when comparing to \eqref{JNWMY}.

While JNW does not admit KS coordinates, we can express it in a similar manner, inspired by the generalised KS form of double field theory  \cite{Lee:2018gxc}. We start by defining the area-radius coordinate
\begin{align}
R^2=e^{2\phi}\left(1-\frac{r_0}{r}\right)^{\frac{-a}{r_0}}r(r-r_0)~,
\end{align}
such that the metric reads
\begin{equation}
\begin{gathered}
\df s^2=-f_t(r)\,\df t^2
+f_R(r)\, \df R^2+R^2\,\df \Omega_2^2~,\\
f_t(r)=\left(1-\frac{r_0}{r}\right)^\frac{a+b}{r_0}~,\qquad
f_R(r)=\frac{4 r (r-r_0)}{(2\,r-a+b-r_0)^2}~.
\end{gathered}
\end{equation}
Changing to ingoing  Eddington-Finkelstein coordinates,
\begin{align}
\df &v=\df t +\sqrt{\frac{f_R(r)}{f_t(r)}}\,\df R~,\nonumber\\
\df s^2&=-\df v^2+2\sqrt{f_t(r)\,f_R(r)}\,\df v\,\df R+R^2\,\df \Omega^2_2~,\nonumber\\
&=-\df v^2+2\df v\df R+R^2\,\df \Omega^2_2+(1-f_t(r))\df v\left(\df v+\frac{2\sqrt{f_t(r)\,f_R(r)}}{1-f_t(r)}\df R\right)~,
\end{align}
where the first three terms are the flat background metric. Let us define two auxiliary variables and another change of coordinates:
\begin{align}
v&=T+R~,\nonumber\\
V&\equiv 1-f_t(r)=1-\left(1-\frac{r_0}{r}\right)^\frac{a+b}{r_0}~,\\
\Omega&\equiv 1-\frac{2}{V}(1-\sqrt{f_t(r)f_R(r)})        \nonumber\\
&=1-\frac{2}{V}\left[1-\left(1-\frac{r_0}{r}\right)^\frac{r_0+a+b}{r_0}\left(1-\frac{r_0+a-b}{2r}\right)^{-1}\right]~.
\end{align}
The line element is transformed into
\begin{equation}
\begin{gathered}
\df s^2=-\df T^2+\df R^2+R^2\,\df \Omega^2_2+V\,l\,\bar{l}~,\\
l=\df T+\df R~,\qquad
\bar{l}=\df T+\Omega\,\df R~,
\end{gathered}
\end{equation}
which is reminiscent of KS. Note, however, that only in the Schwarzschild case (i.e., $b= 0 $, $\Omega=1$) do $l$ and $\bar{l}$ coincide, and we recover the standard KS form of Schwarschild. Moreover, the metric does not even admit the DFT generalisation of the KS metric \cite{Lee:2018gxc}, because $\bar{l}$ is not null unless $a=0$ or $b=0$. A relaxation of the DFT KS form is threfore required. Some of the properties of the vectors still hold:
\begin{align}
l^\mu l_\mu=0~,\qquad \bar{l}^\mu\partial_\mu l_\nu=0~,\qquad\bar{l}^\mu\partial_\nu l_\mu=0~,\label{properties_l}\\
l^\mu\bar{l}_\mu\neq 0 ~,\qquad
 l^\mu\partial_\mu l_\nu=0~,\qquad l^\mu\partial_\nu l_\mu=0~,
\end{align}
where we have used the flat metric to contract indices. Finally, we can express it in Cartesian coordinates,
\begin{equation}
\begin{gathered}
\df s^2=-\df T^2+\df X_i \,\df X_i+V\,l\,\bar{l}~,\\
l=\df T+\frac{X_i}{R}\,\df X_i~,\qquad
\bar{l}=\df T+\Omega\,\frac{X_i}{R}\,\df X_i~.
\end{gathered}
\label{KS_like_form}
\end{equation}
Setting
\begin{equation}
\varphi=-V\,\left(1-\frac{r_0}{r}\right)^{-\frac{r_0+a+b}{2r_0}}\left(1-\frac{r_0+a-b}{2\,r}\right)~,
\end{equation}
the metric can be written in the form
\begin{equation}
\df s^2=-\df T^2+\df X_i\,\df X_i-\frac{\varphi}{1+\frac{\varphi}{2}(l\cdot\bar{l})}\,l\,\bar{l}~,
\label{KS_like_form2}
\end{equation}
which obeys the KS-like ansatz \eqref{the_ansatz} to be used in section \ref{eq:DFTrelax}.
In this coordinate system the following relations also hold
\begin{align}
\det g&=-\left(\frac{V(1-\Omega)}{2}-1\right)^2~,\\
\varphi&=-\frac{V}{\sqrt{-\det g}}~,\\[0.5em]
\Omega&=1-2\,(V^{-1}+\varphi^{-1})~.
\end{align}

\subsection{Double field theory and the relaxed Kerr-Schild ansatz}
\label{eq:DFTrelax}

Double field theory (DFT) is a closed string effective field theory with manifest T-duality, where the latter is expressed by $\mathit{O}(D,D)$ covariance in a `doubled spacetime' where points are labelled as $(x^\mu,\tilde x_\mu)$. It provides a unified geometric framework for the entire massless NS-NS sector, encoded in an $\mathit{O}(D,D)$ covariant manner in the DFT fields, which are the generalised metric $\mathcal{H}_{MN}$ and the DFT dilaton $d$. Here, $M,N, \cdots=1,\cdots,2D$ are $\mathit{O}(D,D)$ vector indices. The generalised metric is a symmetric rank-2 $\mathit{O}(D,D)$ tensor satisfying the $\mathit{O}(D,D)$ constraint,
\begin{equation}
\mathcal{H}_{MP} \mathcal{J}^{PQ} \mathcal{H}_{QN}= \mathcal{J}_{MN}\,,
\label{OddConstraint}\end{equation}
where $\mathcal{J}_{MN}$ is the $\mathit{O}(D,D)$ metric 
\begin{equation}
  \mathcal{J}_{MN} = \begin{pmatrix} 0 & \delta^{\mu}{}_{\nu} \\ \delta_{\mu}{}^{\nu}& 0 \end{pmatrix}\,, \qquad \mathcal{J}^{MN} = \begin{pmatrix} 0 & \delta_{\mu}{}^{\nu} \\ \delta^{\mu}{}_{\nu}& 0 \end{pmatrix}\,,
\label{}\end{equation}
which defines the inner product and raises and lowers the $\mathit{O}(D,D)$ vector indices. One can solve the $\mathit{O}(D,D)$ constraint such that the generalised metric $\mathcal{H}$ and the DFT dilaton $d$ encode the usual string-frame massless NS-NS fields as follows:
\begin{equation}
 \mathcal{H}_{MN} = \begin{pmatrix} g^{\mu\nu} & - g^{\mu\rho} B_{\rho \nu} \\ B_{\mu \rho} g^{\rho\nu} & g_{\mu\nu}-B_{\mu\rho}g^{\rho\sigma}B_{\sigma\nu}\end{pmatrix}\,,\qquad e^{-2d}=\sqrt{-g}e^{-2\phi}\,.
\label{parametrisations}\end{equation}
In general, $\mathit{O}(D,D)$ vectors unify a $D$-dimensional vector and form field pair into a single object. For example, an arbitrary $\mathit{O}(D,D)$ vector $V_{M}$ is parametrised in terms of a $D$-dimensional vector $v^{\mu}$ and a form field $k_{\mu}$ as
\begin{equation}
  V_{M} = \begin{pmatrix} v^{\mu}\\k_{\mu}\end{pmatrix}\,, \qquad \mathrm{and}\qquad V^{M} = \mathcal{J}^{MN} V_{N} = \begin{pmatrix} k_{\mu} \\ v^{\mu}\end{pmatrix}\,.
\label{}\end{equation}

An important feature of DFT related to the double copy is the doubled local Lorentz group, $\mathit{O}(1,D-1)_{L}\times \mathit{O}(1,D-1)_{R}$, which is the maximally compact subgroup of $\mathit{O}(D,D)$ including the Lorentz group. The doubled local Lorentz group originates in the left-right mode decomposition of the closed string, and shares the same origin as the KLT relations \cite{Kawai:1985xq} in string scattering amplitudes, which underlie the double copy. This structure is transparent if we introduce a chiral and anti-chiral basis in the doubled vector space. One may recast the $\mathit{O}(D,D)$ constraint as $\mathcal{H}_{M}{}^{P} \mathcal{H}_{P}{}^{N}= \delta_{M}{}^{N}$, and it defines a pair of projection operators,
\begin{equation}
  P_{M}{}^{N} = \frac{1}{2} \big(\delta_{M}{}^{N} + \mathcal{H}_{M}{}^{N}\big)\,, \qquad \bar{P}_{M}{}^{N} = \frac{1}{2} \big(\delta_{M}{}^{N} - \mathcal{H}_{M}{}^{N}\big)\,.
\label{projectors}\end{equation}
These project the doubled vector space into chiral and anti-chiral sectors which correspond to the left- and right-moving sectors, respectively. 

Motivated by the KS-like form of the JNW metric \eqref{KS_like_form}, we introduce an ansatz for $\mathcal{H}$ and $d$ in terms of two $\mathit{O}(D,D)$ vectors, $K_{M}$ and $\bar{K}_{M}$, where $K$ is null but $\bar{K}$ does not have to be null in general,
\begin{equation}
  K_{M}K^{M} = 0\,, \qquad \bar{K}_{M} \bar{K}^{M} \neq 0\,.
\label{relaxed_null}\end{equation}
Let us consider a flat background, $g_{0\mu\nu} = \eta_{\mu\nu}$, $B_{\mu\nu}=0$ and $\phi=\mathrm{constant}$, and denote the corresponding background DFT fields as $\mathcal{H}_{0}$ and $d_{0}$, where
\begin{equation}
  \mathcal{H}_{0MN}=\begin{pmatrix} \eta^{\mu\nu}&0\\0&\eta_{\mu\nu}\end{pmatrix} \,,\qquad d_{0} = \mathrm{constant}\,.
\label{paraH0}\end{equation}
We associate to $\mathcal{H}_{0}$ a pair of background projection operators $P_{0}$ and $\bar{P}_{0}$ via \eqref{projectors}. As we have described above, the chiralities are closely related to the underlying structure of the double copy, hence we require definite chiralities on $K_{M}$ and $\bar{K}_{M}$ for the manifest left and right mode decomposition,
\begin{equation}
  P_{0M}{}^{N} K_{N} = K_{M}\,, \qquad \bar{P}_{0M}{}^{N} \bar{K}_{N} = \bar{K}_{M}\,.
\label{}\end{equation}
This implies that $K$ and $\bar{K}$ are orthogonal, $K_{M} \bar{K}^{M} = 0$. One may solve the above chirality conditions explicitly using \eqref{paraH0}, which yields
\begin{equation}
  K_{M} = \frac{1}{\sqrt{2}} \begin{pmatrix} l^{\mu} \\ \eta_{\mu\nu}l^{\nu}\end{pmatrix}\,, \qquad \bar{K}_{M} = \frac{1}{\sqrt{2}} \begin{pmatrix} \bar{l}^{\mu} \\ -\eta_{\mu\nu}\bar{l}^{\nu}\end{pmatrix}\,.
\label{ParaK}\end{equation}

Now we are ready to write down a KS-like ansatz for the generalised metric:
\begin{equation}
\begin{aligned}
    \mathcal{H}_{MN} &= \mathcal{H}_{0 MN} +  \kappa \varphi \big(K_{M}\bar{K}_{N} + K_{N}\bar{K}_{M}\big) - \frac{\kappa^{2}}{2}\varphi^{2} \bar{K}^{2} K_{M} K_{N}\,,
  \\
  d &= d_{0} + \kappa f\,,
\end{aligned}\label{KS_like_ansatz}
\end{equation}
where $\kappa$ is an expansion parameter. We refer to this form as the `relaxed KS ansatz' because the null condition for the DFT KS ansatz of \cite{Lee:2018gxc} is partially relaxed; the latter is recovered when $\bar{K}$ is a null vector. Though the null condition is relaxed, the new ansatz satisfies the $\mathit{O}(D,D)$ constraint \eqref{OddConstraint} automatically without further truncation. Substituting the parametrisation of $K$ and $\bar{K}$ in \eqref{ParaK} into \eqref{relaxed_null}, we obtain conditions on $l$ and $\bar{l}$:
\begin{equation}
  l_{\mu} l^{\mu} =0 \,,\qquad \bar{l}_{\mu} \bar{l}^{\mu}\neq 0\,, \qquad l_{\mu} \bar{l}^{\mu}\neq 0\,,
\label{}\end{equation}
which are consistent with the JNW geometry as expressed in \eqref{KS_like_form}. Interestingly, the feature of the partially relaxed null condition is analogous to previous studies such as the `extended' KS ansatz \cite{Ett:2010by} and the heterotic KS ansatz \cite{Cho:2019ype}. From the parametrisation of $\mathcal{H}$, we can easily read off the corresponding ansatz for the metric and Kalb-Ramond field:
\begin{equation}
\begin{aligned}
  g_{\mu\nu} &= \eta_{\mu\nu} - \frac{\kappa\varphi}{1+\frac{\kappa\varphi}{2}(l\cdot \bar{l})} l_{(\mu} \bar{l}_{\nu)} \,,
  \\
  g^{\mu\nu} &= \eta^{\mu\nu} + \kappa\varphi l^{(\mu} \bar{l}^{\nu)} + \frac{\kappa^{2}\varphi^{2} \bar{l}^{2}}{4}  l^{\mu} l^{\nu}\,,
  \\
  B_{\mu\nu} &= \frac{\kappa\varphi}{1+\frac{\kappa\varphi}{2}(l\cdot \bar{l})} l_{[\mu} \bar{l}_{\nu]}\,.
\end{aligned}\label{the_ansatz}
\end{equation}
It can easily be seen that the JNW solution fits this ansatz. 
The JNW metric was written precisely in this form in \eqref{KS_like_form2}; we kept $\kappa\neq 1$ here for clarity. As for the Kalb-Ramond field, given the JNW expressions for $\varphi$, $l$ and $\bar{l}$, it is of the form $B=B(r)\,\df R\,\wedge\df T$. Since $r$ is a function of $R$ only, $B_{\mu\nu}$ is pure gauge and it can be set to zero.

\subsection{DFT equations of motion and the single copy}

The field equations of DFT are given by the generalised curvatures, analogously to general relativity.\footnote{See appendix~\ref{sec:appendix} for a concise review of the equations of motion in DFT.}  The generalised curvature scalar $\mathcal{R}$ and tensor $\mathcal{R}_{MN}$ defined in \eqref{genCurvature} are the equations of motion of the DFT dilaton and the generalised metric, respectively,
\begin{equation}
 \mathcal{R}=0\,, \qquad \mathcal{R}_{\mu\nu}=0\,,
\label{}\end{equation}
where $\mathcal{R}_{\mu\nu}$ is a pullback of $\mathcal{R}_{MN}$ into the $D$-dimensional spacetime. Note that $\mathcal{R}_{\mu\nu}$ is not symmetric nor antisymmetric: the symmetric and antisymmetric parts are the equations of motion for the metric and the Kalb-Ramond field, respectively. These reproduce the supergravity equations of motion for the massless NS-NS fields in the string frame,
\begin{align}
R_{\mu\nu}+2\nabla_\mu\nabla_\nu\phi-\frac{1}{4}H_{\mu\rho\sigma}H_\nu^{\ \ \rho\sigma}=0~,\nonumber \\ 
R+4\,\square\phi-4\,\nabla^\mu\phi\nabla_\mu\phi-\frac{1}{12}H_{\mu\nu\rho}H^{\mu\nu\sigma}=0~,\label{sugra_eom} \\
\nabla^\rho H_{\rho\mu\nu} -2H_{\rho\mu\nu}\nabla^\rho\phi=0~,\nonumber
\end{align}
which follow from the action \eqref{eq:stringframeaction}.

Let us now discuss the field equations subject to the relaxed KS ansatz \eqref{KS_like_ansatz}. Recall that, in the KS ansatz, an additional constraint is required in order to linearise the equations of motion, which in the case of general relativity is the geodesic condition on the null vector field. Such a constraint is obtained by contracting the null vectors with the free indices of the (generalised) curvature tensor. In the case of our relaxed KS ansatz, however, it is very cumbersome to work with this constraint. Therefore, we will assume a stronger constraint, which is satisfied in a class of solutions that includes JNW. We impose
\begin{equation}
  \bar{K}^{M} \partial_{M} K_{N} = 0\,, 
\label{constraint}\end{equation}
which reduces to the second equation of \eqref{properties_l}. Note that, while the analogous condition also appeared in the KS ansatz of DFT \cite{Lee:2018gxc}, in our ansatz we allow for $K^{M} \partial_{M} \bar{K}_{N} \neq 0$. Given the null condition on $l$ and the constraint \eqref{constraint}, the generalised curvature tensor reduces to
\begin{equation}
\begin{aligned}
   \mathcal{R}_{\mu\nu} &= \frac{1}{4} e^{2\kappa f} \partial^{\rho}\Big[\,e^{-2\kappa f} \Big(\partial_{\rho}\big(\kappa \varphi l_{\mu} \bar{l}_{\nu} \big) - \partial_{\mu} \big( \kappa\varphi l_{\rho} \bar{l}_{\nu} \big) - \partial_{\nu} \left( \kappa\varphi l_{\mu} \bar{l}_{\rho} \right) -\kappa^{2}\varphi^{2} l_{\mu}l^{\sigma} \bar{l}_{[\nu}\partial_{|\sigma|}\bar{l}_{\rho]}\Big)\, \Big] 
  \\ 
  &\quad + \kappa\partial_{\mu} \partial_{\nu}f +\frac{\kappa^{2}}{2} \Big(\, \varphi l^{\rho} \bar{l}_{\nu} \partial_{\mu}\partial_{\rho} f  +\varphi l_{\mu} \bar{l}^{\rho} \partial_{\nu}\partial_{\rho} f\, \Big)-\frac{\kappa^{2}}{8} \varphi^{2}\bar{l}^{2} \partial_{\mu}l_{\rho}\partial_{\nu}l^{\rho}
  \\
  &\quad +\frac{\kappa^{2}}{4} \partial^{\rho} \big( \varphi^{2}\bar{l}^{2} l_{[\mu} \partial_{|\nu|} l_{\rho]} \big) -\frac{\kappa^{2}}{8} \varphi l_{\mu} \bar{l}^{\sigma} \partial_{\nu}\partial_{\rho} (\varphi l^{\rho} \bar{l}_{\sigma})  + \frac{\kappa^{3}}{4}\varphi l^{\rho} \bar{l}^{\sigma} \partial_{\nu}\big(\varphi l_{\mu} \bar{l}_{\sigma}\partial_{\rho}f\big)= 0 \,.
\end{aligned}\label{}
\end{equation}

We now discuss how to extract the single copy from $\mathcal{R}_{\mu\nu}$. By carrying out the same procedure described in the conventional KS formalism, it can be shown that one obtains the Maxwell equations from the gravity equations of motion. Suppose that the relaxed KS geometry admits at least one Killing vector $\xi$. We also assume that the Killing vector is constant in our choice of coordinates, and satisfies $\xi^{\nu} \partial_{\nu} \mathcal{F}_{\mu_{1}\cdots \mu_{n}}=0$, where $\mathcal{F}_{\mu_{1}\cdots \mu_{n}}$ is an arbitrary tensor field. We will be interested in the timelike Killing vector $\xi=\partial_T$ for JNW. The single copy can be realised by contracting the Killing vector $\xi$ with one of the free indices of the field equations of the generalised metric, $\mathcal{R}_{\mu\nu}$. We further require that $l$, $\bar{l}$ and $\xi$ are normalised as $\xi\cdot l = \xi\cdot \bar{l} = 1$, which is directly the case for JNW in \eqref{KS_like_form}. Such a normalisation is always possible, since the KS form is preserved under the rescaling of $l$ and $\bar{l}$. Recall that $\mathcal{R}_{\mu\nu}$ is not symmetric nor antisymmetric, thus there are two distinct equations:
\begin{equation}
\begin{aligned}
  \xi^{\nu} \mathcal{R}_{\mu\nu} &= \frac{1}{4}e^{2f} \partial^{\rho} \Big[\, 2\partial_{[\rho} (\tilde{\varphi} l_{\mu]}) + 4\tilde{\varphi} l_{[\mu}\partial_{\rho]}f - \frac{1}{2}e^{2f}\tilde{\varphi}^{2}l^{\sigma} l_{\mu}\partial_{\sigma}\bar{l}_{\rho}\big)\,\Big] +\frac{1}{2} e^{2f}\tilde{\varphi} l^{\rho} \partial_{\rho}\partial_{\mu}f \,,
  \\
  \xi^{\mu} \mathcal{R}_{\mu\nu} &= \frac{1}{4} e^{2f}\Big[\, \partial^{\rho}\big(2\partial_{[\rho} (\tilde{\varphi} \bar{l}_{\nu]}) + 4 \tilde{\varphi} \bar{l}_{[\nu}\partial_{\rho]}f - e^{2f}\tilde{\varphi}^{2} l^{\sigma} \bar{l}_{[\nu}\partial_{|\sigma|}\bar{l}_{\rho]}\big) +2 \tilde{\varphi} \bar{l}^{\sigma} \partial_{\sigma}\partial_{\nu}f
  \\
  &\quad +\frac{1}{2} \tilde{\varphi} \bar{l}^{\sigma} \big(\partial_{\rho}(e^{2f}\tilde{\varphi} \bar{l}_{\sigma})\partial_{\nu}l^{\rho} -\partial_{\rho}(\partial_{\nu}(e^{2f} \tilde{\varphi}\bar{l}_{\sigma}) l^{\rho}) + 2l^{\rho} \partial_{\nu}(e^{2f}\tilde{\varphi}\bar{l}_{\sigma}\partial_{\rho}f)\big)\Big] \,,
\end{aligned}\label{single_copy1}
\end{equation}
where we defined $\tilde{\varphi} = e^{-2f} \varphi$ and we set $\kappa=1$ for simplicity.

It is not immediately obvious how to extract the single copy from \eqref{single_copy1} due to the higher order terms in $\kappa$, as opposed to the simpler case of the DFT KS ansatz. However, the terms linear in $\kappa$ in overlap with the analogous computation in the DFT KS case. Thus one may guess that the higher order terms would be extra contributions over the KS single copy relation, where the two gauge fields are proportional to $l_\mu$ and $\bar l_\mu$. Let us collect the higher order terms, and express them with the help of a pair of auxiliary vector fields $C_{\mu}$ and $\bar{C}_{\mu}$, obeying
\begin{equation}
\begin{aligned}
  \partial^{\rho}\partial_{[\rho}C_{\mu]} &= -\partial^{\rho}\Big( \frac{1}{4}e^{2f}  \tilde{\varphi}^{2} l_{\mu} l^{\sigma} \partial_{\sigma}\bar{l}_{\rho} -2\tilde{\varphi} l_{[\mu}\partial_{\rho]}f \Big)  + \tilde{\varphi} l^{\rho} \partial_{\rho} \partial_{\mu} f\,,
  \\
  \partial^{\rho}\partial_{[\rho} \bar{C}_{\mu]} &= - \partial^{\rho} \Big( \frac{1}{2}e^{2f}\tilde{\varphi}^{2} l^{\sigma} \bar{l}_{[\nu}\partial_{|\sigma|}\bar{l}_{\rho]} -2\tilde{\varphi}\bar{l}_{[\nu}\partial_{\rho]}f \Big) +\tilde{\varphi} \bar{l}^{\sigma} \partial_{\sigma}\partial_{\nu}f
  \\ 
  &\quad +\frac{1}{4} \tilde{\varphi} \bar{l}^{\sigma} \Big(\partial_{\rho}(e^{2f}\tilde{\varphi} \bar{l}_{\sigma})\partial_{\nu}l^{\rho} -\partial_{\rho}\big(\partial_{\nu}(e^{2f} \tilde{\varphi}\bar{l}_{\sigma}) l^{\rho}\big) + 2l^{\rho} \partial_{\nu}(e^{2f}\tilde{\varphi}\bar{l}_{\sigma}\partial_{\rho}f)\Big)\,.
\end{aligned}\label{conditionw_f}
\end{equation}
Notice that these definitions are possible because the currents on the right-hand side are conserved, by virtue of the equations of motion, i.e., $\partial^\mu\partial^{\rho}\partial_{[\rho}C_{\mu]} =  \partial^\mu\partial^{\rho}\partial_{[\rho} \bar{C}_{\mu]} =0$. Even though the equations look rather complicated to solve, the Killing direction components can be easily integrated as
\begin{equation}
\begin{aligned}
  \partial^{\rho}\Big(\partial_{\rho}C_{\xi} + \frac{1}{2}e^{2f}  \tilde{\varphi}^{2} l^{\sigma} \partial_{\sigma}\bar{l}_{\rho} -2\tilde{\varphi} \partial_{\rho}f \Big) &= 0 \,,
  \\
  \partial^{\rho}\Big(\partial_{\rho} \bar{C}_{\xi}+\frac{1}{2}e^{2f}\tilde{\varphi}^{2} l^{\sigma} \partial_{\sigma}\bar{l}_{\rho} -2\tilde{\varphi} \partial_{\rho}f\Big) &= 0\,,
\end{aligned}\label{Killing_direction}
\end{equation}
where $C_{\xi} = \xi^{\mu} C_{\mu}$ and $\bar{C}_{\xi} = \xi^{\mu} \bar{C}_{\mu}$, and we have used the normalisation, $l_{\xi}=\bar{l}_{\xi}=1$. This indicates that $C_{\xi}$ and $\bar{C}_{\xi}$ should be identified; indeed, that will be required by the uniqueness of the `zeroth' copy to be discussed shortly. 
As for the other components of $C$ and $\bar{C}$, we have to treat them case by case. We will discuss the JNW example in the next subsection.

 Making the use of the auxiliary fields, \eqref{single_copy1} reduces to the following compact form,
\begin{equation}
\begin{aligned}
  4 e^{-2f}\xi^{\nu} \mathcal{R}_{\mu\nu} &= \partial^{\rho}\Big[\, \partial_{\rho}(\tilde{\varphi} l_{\mu}+ C_{\mu})-\partial_{\mu}(\tilde{\varphi} l_{\rho} + C_{\rho})\,\Big]=0\,,
  \\
  4 e^{-2f}\xi^{\mu} \mathcal{R}_{\mu\nu} &= \partial^{\rho} \Big[\,\partial_{\rho}(\tilde{\varphi} \bar{l}_{\nu}+\bar{C}_{\nu})
-\partial_{\nu}(\tilde{\varphi} \bar{l}_{\rho}+\bar{C}_{\rho})\,\Big]=0\,.
\end{aligned}\label{single_copy2}
\end{equation}
This can be interpreted as a pair of Maxwell equations 
\begin{equation}
\begin{aligned}
  \partial^{\mu} F_{\mu\nu} = 0\,, \qquad \partial^{\mu} \bar{F}_{\mu\nu}= 0\,,
\end{aligned}
\end{equation}
by identifying the gauge fields as the single copy
\begin{equation}
  A_{\mu} = \tilde{\varphi} l_{\mu}+ C_{\mu}\,, \qquad \bar{A}_{\mu} = \tilde{\varphi} \bar{l}_{\mu}+ \bar{C}_{\mu}\,.
\label{single_copy}\end{equation}
Here, $F_{\mu\nu}$ and $\bar{F}_{\mu\nu}$ are the field strengths of the $A_{\mu}$ and $\bar{A}_{\mu}$ respectively. This ensures that solutions of \eqref{sugra_eom} with the form of \eqref{the_ansatz}, subject to the constraint \eqref{constraint}, can be represented by a pair of Maxwell gauge fields. We emphasise again that $A_{\mu}$ and $\bar{A}_{\mu}$ are associated with left and right movers in a string theory interpretation, which is consistent with the double copy.

Finally, we can consider also the `zeroth copy'. This is the scalar analogue of the gravity and gauge-theory solutions. Since the gauge-theory solutions are Abelian, we expect that the scalar will also be an `Abelianised' version of the bi-adjoint scalar field $\Phi^{aa'}$. It is obtained in our formalism by contracting the Killing vector into both free indices of $\mathcal{R}_{\mu\nu}$, leading to a scalar equation of motion. One may use the result of \eqref{single_copy2} to get a pair of d'Alembertian equations,
\begin{equation}
  \Box (\tilde{\varphi} +C_{\xi}) = 0\,,\qquad   \Box (\tilde{\varphi} +\bar{C}_{\xi}) = 0\,.
\label{}\end{equation}
As we mentioned, $C_{\xi}$ and $\bar{C}_{\xi}$ should be identified. The single copy can therefore be recognised as
\begin{equation}
  \Phi = \tilde{\varphi} +C_{\xi} = \tilde{\varphi} + \bar{C}_{\xi}\,.
\label{zeroth_copy}\end{equation}
%

\subsection{JNW and Coloumb}
So far we have considered a general construction of the single copy for the relaxed KS ansatz \eqref{the_ansatz}. We now apply the previous formalism to the JNW case, and show that the corresponding single copy is the Coulomb potential (i.e., both $A_\mu$ and $\bar A_\mu$ are Coulomb). As noted before, we need to determine the auxiliary vector fields $C_{\mu}$ and $\bar{C}_{\mu}$ to spell out the single copy. Since the JNW geometry is static, with timelike Killing vector $\xi=\partial_T$, $C_{T}$ and $\bar{C}_{T}$ can be solved straightforwardly from \eqref{Killing_direction}. If we substitute all the necessary data, we get 
\begin{equation}
\begin{aligned}
  \partial_{r}C_{T}(r) &= \partial_{r}\bar{C}_{T}(r) = -e^{-2f} \Big(\frac{\varphi^{2}}{V^{2}}\partial_{r}V+\partial_{r}\varphi+2 e^{-2f} \partial_{r}f \Big)\,
\end{aligned}\label{}
\end{equation}
in the asymptotically decaying case.
The field strengths associated to \eqref{single_copy} satisfy
\begin{equation}
  F_{iT} =\bar{F}_{iT} = (a+b) r^{-2}\Big(1-\frac{r_{0}}{r}\Big)^{\frac{-r_{0}+a-b}{r_{0}}} l_{i}= \frac{a+b}{R^{2}}\,l_{i} = \frac{(a+b)}{R^{3}}\,X_{i}\,.
\label{}\end{equation}
These are nothing but the electric field for the Coulomb potential, and it turns out that all other components of the field strengths vanish. In particular, we can easily show that the spatial components of the static, spherically symmetric gauge fields $A_{\mu}$ and $\bar{A}_{\mu}$ are pure gauge. This is better seen in spherical coordinates, where the only non-vanishing spatial component of $l$, $\bar{l}$, $C$ or $\bar{C}$ is the radial one, and it only depends on the radial coordinate, which is also the case for $\tilde\varphi$. Therefore, the relevant spatial vector fields are all curl free, %
\begin{equation}
  \partial_{[i} (\tilde{\varphi} l_{j]}) = \partial_{[i} (\tilde{\varphi} \bar{l}_{j]}) = \partial_{[i} C_{j]} = \partial_{[i} \bar{C}_{j]} = 0\,.
\label{}\end{equation}
Hence, $A_{i}$ and $\bar{A}_{i}$ are pure gauge, and only $A_{T}$ and $\bar{A}_{T}$ contribute to the field strength.

This shows that the single copy for the JNW solution is given by a point electric charge as expected. The corresponding electric charge parameter is associated to the linear sum of the mass and dilaton coupling in the string frame, $a+b$. As argued earlier in the paper, the two parameters in gravity reduce to one via the single copy.

One interesting point is that the single copy exists and is the same whether the gravity solution is a naked singularity ($b\neq0$) or the Schwarzschild solution ($b=0$). This highlights the fact the single copy does not reflect the causal structure of the gravity solution. Some reflection indicates, however, that this is to be expected. The single copy does not apply to the full metric, but only to the deviation from the Minkowski metric. It is from the interplay between the Minkowski metric and the deviation that the causal structure arises.

Finally, using \eqref{zeroth_copy}, it is straightforward to consider the zeroth copy. As expected, the associated linearised bi-adjoint field for the JNW solution is a Coulombic potential,
\begin{equation}
  \Phi = \tilde{\varphi} + C_{T} = \frac{a+b}{R}\,,
\label{}\end{equation}
which is the static, spherically symmetric solution that decays asymptotically.

Therefore, both the single and zeroth copies for the JNW solution coincide with those of the Schwarzschild solution, up to irrelevant constant factors.


\section{Conclusion}
\label{sec:conclusion}

In this paper, we claim that the most general double copy of the Coulomb solution is the JNW solution, which includes a mass parameter and a dilaton parameter. The JNW solution reduces to the Schwarzschild solution if the dilaton vanishes, which is consistent with previous work in the vacuum case. We support our claim both in perturbation theory, extending the `fat graviton' analysis of \cite{Luna:2016hge}, and in an exact map, extending the double field theory Kerr-Schild ansatz of \cite{Lee:2018gxc}. One remarkable feature of the latter approach is that it exhibits the double copy origin of Kerr-Schild-type maps of solutions between gravity and gauge theory, by associating the pair of Kerr-Schild-type vectors to left and right movers in closed string theory. Moreover, it shows that, when the dilaton is turned on, the exact double copy is best expressed in the string frame, rather than the Einstein frame.

There are several directions of interest for future work. One unsatisfying aspect of our work is that the perturbative approach and the exact approach discussed here were not explicitly connected. Due to gauge choices and field redefinitions, this appears to be cumbersome. And yet it would be very interesting to see how the perturbative double copy could be resummed into the exact double copy. This would provide important clues as to what is the exact double copy analogue of the colour-kinematics duality, which underlies the double copy in scattering amplitudes. 

On the exact double copy front, an extension of the analysis in this paper would allow us to study the double copy interpretation of the most general static, spherically symmetric and asymptotically flat solution to NS-NS gravity, which is known \cite{Burgess:1994kq}. It is more general than the JNW solution, in that it admits a B-field whose field strength is spherically symmetric. The `single copy' is not, however, the Coulomb solution, since two distinct gauge-theory solutions are required in order to introduce the antisymmetric B-field via the double copy. The extension of our work to heterotic double field theory, building on \cite{Cho:2019ype}, would be interesting too.

It should also be possible to extend the vacuum Weyl double copy \cite{Luna:2018dpt}, and a higher-dimensional version based on \cite{Monteiro:2018xev}, to include the dilaton and B-field, using ideas from double field theory. In fact, this could potentially elucidate some aspects of the generalised curvatures of double field theory, alluded to in the appendix. 

It is clear that there is much more to explore in how double field theory expresses the double copy.


\section*{Acknowledgements}
We would like to thank Andres Luna, Alex Ochirov, Donal O'Connell and Chris White for discussions. KK and KL are supported by IBS under the project code, IBS-R018-D1. RM and DPV are supported by the Royal Society, through a University Research Fellowship and a studentship, respectively. 

\appendix

\section{Equations of motion in double field theory}
\label{sec:appendix}

%
%

We review here the derivation of the double field theory (DFT) equations of motion. The covariant derivative and its curvature tensors are defined with respect to the so-called generalised Lie derivative or generalised diffeomorphism. It plays the role of gauge symmetry in DFT, and acts on the DFT field content as 
\begin{equation}
\begin{aligned}
  (\hat{\cal L}_{X} \mathcal{H})_{MN} &= X^{P} \partial_{P} \mathcal{H}_{MN} + (\partial_{M} X^{P} - \partial^{P} X_{M}) \mathcal{H}_{PN} + (\partial_{N} X^{P} - \partial^{P} X_{N}) \mathcal{H}_{MP}\,,
\\
\hat{\cal L}_{X} d &= X^{M} \partial_{M} d - \frac{1}{2} \partial_{M} X^{M}\,,
\end{aligned}\label{genLie}
\end{equation}
where, with respect to the generalised Lie derivative, the generalised metric $\mathcal{H}$ is a rank-2 tensor and the DFT dilaton $d$ is a scalar density. The gauge parameter $X^{M}$ combines the diffeomorphism parameter $\xi^{\mu}$ and the one-form gauge parameter $\Lambda_{\mu}$ for the Kalb-Ramond field in an $\mathit{O}(D,D)$ covariant manner
\begin{equation}
  X^{M} = \{\xi^{\mu}\,, \Lambda_{\mu}\}\,.
\label{}\end{equation}
For closure of the algebra of generalised diffeomorphisms (i.e., the Jacobi identity for $\hat{\mathcal{L}}$), we have to impose the section condition
\begin{equation}
  \partial_{M} \partial^{M} \mathcal{F}_{1} = 0 \,, \qquad \partial_{M}\mathcal{F}_{1} \partial^{M} \mathcal{F}_{2} = 0\,,
\label{}\end{equation}
where $\mathcal{F}_{1}$ and $\mathcal{F}_{2}$ are arbitrary functions on doubled space. The section condition is equivalent to ignoring the winding coordinate $\tilde{x}$ dependence,
\begin{equation}
  \partial_{M}=\left(\begin{array}{l}{\tilde{\partial}^{\mu}} \\ {\partial_{\mu}}\end{array}\right)=\left(\begin{array}{l}{0} \\ {\partial_{\mu}}\end{array}\right)\,.
\label{}\end{equation}

As for the covariant differential operator of the generalised Lie derivative (\ref{genLie}), we define the covariant derivative acting on an $\mathit{O}(D,D)$ tensor as
\begin{equation}
  {\mathcal D}_{M} T_{N_{1}N_{2}\cdots N_{n}} = \partial_{M} T_{N_{1}N_{2}\cdots N_{n}} + \sum_{i=1}^{n}\Gamma_{M N_{i}}{}^{P} T_{N_{1}\cdots P \cdots N_{n}} \,, 
\label{MasterDerivative}\end{equation}
where $\Gamma_{MNP}$ is the DFT connection \cite{Jeon:2011cn,Jeon:2010rw}. One may try to obtain the DFT connection using the compatibility and torsion-free conditions, analogously to Riemannian geometry. However, it turns out that these conditions are not sufficient for determining all the components. Fortunately, one can project out the undetermined part using the projection operators \eqref{projectors}, and the determined part is 
\begin{equation}
\begin{aligned}
  \Gamma_{PMN} = & 2(P\partial_{P} P \bar{P} )_{[MN]} 
	+ 2 (\bar{P}_{[M}{}^{Q} \bar{P}_{N]}{}^{R} 
	- P_{[M}{}^{Q} P_{N]}{}^{R} ) \partial_{Q} P_{R P} 
  \\
  & - \frac{4}{D-1} \big(\bar{P}_{P[M} \bar{P}_{N]}{}^{Q} + P_{P[M} P_{N]}{}^{Q}) 
		\big(\partial_{Q}d + (P\partial^{R} P \bar{P})_{[RQ]}\big)\,.
\label{conn}\end{aligned}\end{equation}

Let us turn to the curvature tensors $\mathcal{R}$ and $\mathcal{R}_{MN}$ in terms of the DFT connection \eqref{conn}. First, we introduce 4-index object $S_{MNPQ}$ defined as
\begin{equation}
  S_{MNPQ} = \frac{1}{2} \big(R_{MNPQ} + R_{PQMN} - \Gamma^{R}{}_{MN} \Gamma_{RPQ} \big)\,,
\end{equation}
where $R_{MNPQ}$ is defined from the standard commutator of the covariant derivatives
\begin{equation}
  R_{MNPQ} = \partial_{M}\Gamma_{NPQ} - \partial_{N}\Gamma_{MPQ} + \Gamma_{MP}{}^{R} \Gamma_{NRQ} -  \Gamma_{NP}{}^{R} \Gamma_{MRQ} \,.
\label{}\end{equation}
One can show that $S_{MNPQ}$ satisfies symmetry properties analogous to the Riemann tensor, namely $S_{MNPQ}=S_{[MN][PQ]} = S_{[PQ][MN]}$ and the first Bianchi identity,
\begin{equation}
  S_{M[N P Q]}=0\,.
\label{}\end{equation}
However, it is not a tensor with respect to the generalised Lie derivative and cannot be a physically meaningful object. Instead, we can obtain meaningful tensors by contracting $S_{MNPQ}$ with the projection operators. The generalised curvature tensor and scalar are defined as
\begin{equation}
  \mathcal{R}_{MN} = 2P_{(M}{}^{P} \bar{P}_{N)}{}^{Q} P^{RS}S_{RPSQ} \,,\qquad   
  \mathcal{R} = 2 P^{MN} P^{PQ} S_{MPNQ}\,,
\label{genCurvature}\end{equation}
and one can show that these are covariant under $\mathit{O}(D,D)$, as well as under generalised diffeomorphisms. Substituting the parametrisations \eqref{parametrisations}, the equations of motion reduce to the conventional supergravity equations of motion \eqref{sugra_eom}. The generalised curvatures satisfy an identity analogous to that of the Einstein tensor, $\nabla_{\mu} G^{\mu\nu} = 0$, namely \cite{Park:2015bza,Angus:2018mep}
\begin{equation}
\begin{aligned}
   \mathcal{D}_{M}\left(4 P^{M P} \bar{P}^{N Q} \mathcal{R}_{PQ} -\bar{P}^{MN} \mathcal{R}\right)=0\,,
  \qquad
  \mathcal{D}_{M}\left(4 \bar{P}^{MP} P^{NQ} \mathcal{R}_{PQ} +P^{MN} \mathcal{R}\right)=0\,.
\end{aligned}\label{Bianchi2}
\end{equation}

\bibliography{references}

\providecommand{\href}[2]{#2}\begingroup\raggedright\begin{thebibliography}{10}

\bibitem{Kawai:1985xq}
H.~Kawai, D.~C. Lewellen and S.~H.~H. Tye, \emph{{A Relation Between Tree
  Amplitudes of Closed and Open Strings}},
  \href{http://dx.doi.org/10.1016/0550-3213(86)90362-7}{\emph{Nucl. Phys.}
  {\bfseries B269} (1986) 1--23}.

\bibitem{Bern:2008qj}
Z.~Bern, J.~J.~M. Carrasco and H.~Johansson, \emph{{New Relations for
  Gauge-Theory Amplitudes}},
  \href{http://dx.doi.org/10.1103/PhysRevD.78.085011}{\emph{Phys. Rev.}
  {\bfseries D78} (2008) 085011},
  [\href{https://arxiv.org/abs/0805.3993}{{\ttfamily 0805.3993}}].

\bibitem{Bern:2010ue}
Z.~Bern, J.~J.~M. Carrasco and H.~Johansson, \emph{{Perturbative Quantum
  Gravity as a Double Copy of Gauge Theory}},
  \href{http://dx.doi.org/10.1103/PhysRevLett.105.061602}{\emph{Phys. Rev.
  Lett.} {\bfseries 105} (2010) 061602},
  [\href{https://arxiv.org/abs/1004.0476}{{\ttfamily 1004.0476}}].

\bibitem{Bern:2019prr}
Z.~Bern, J.~J. Carrasco, M.~Chiodaroli, H.~Johansson and R.~Roiban, \emph{{The
  Duality Between Color and Kinematics and its Applications}},
  \href{https://arxiv.org/abs/1909.01358}{{\ttfamily 1909.01358}}.

\bibitem{Monteiro:2014cda}
R.~Monteiro, D.~O'Connell and C.~D. White, \emph{{Black holes and the double
  copy}}, \href{http://dx.doi.org/10.1007/JHEP12(2014)056}{\emph{JHEP}
  {\bfseries 12} (2014) 056},
  [\href{https://arxiv.org/abs/1410.0239}{{\ttfamily 1410.0239}}].

\bibitem{Luna:2015paa}
A.~Luna, R.~Monteiro, D.~O'Connell and C.~D. White, \emph{{The classical double
  copy for Taub–NUT spacetime}},
  \href{http://dx.doi.org/10.1016/j.physletb.2015.09.021}{\emph{Phys. Lett.}
  {\bfseries B750} (2015) 272--277},
  [\href{https://arxiv.org/abs/1507.01869}{{\ttfamily 1507.01869}}].

\bibitem{Ridgway:2015fdl}
A.~K. Ridgway and M.~B. Wise, \emph{{Static Spherically Symmetric Kerr-Schild
  Metrics and Implications for the Classical Double Copy}},
  \href{http://dx.doi.org/10.1103/PhysRevD.94.044023}{\emph{Phys. Rev.}
  {\bfseries D94} (2016) 044023},
  [\href{https://arxiv.org/abs/1512.02243}{{\ttfamily 1512.02243}}].

\bibitem{Luna:2016due}
A.~Luna, R.~Monteiro, I.~Nicholson, D.~O'Connell and C.~D. White, \emph{{The
  double copy: Bremsstrahlung and accelerating black holes}},
  \href{http://dx.doi.org/10.1007/JHEP06(2016)023}{\emph{JHEP} {\bfseries 06}
  (2016) 023}, [\href{https://arxiv.org/abs/1603.05737}{{\ttfamily
  1603.05737}}].

\bibitem{White:2016jzc}
C.~D. White, \emph{{Exact solutions for the biadjoint scalar field}},
  \href{http://dx.doi.org/10.1016/j.physletb.2016.10.052}{\emph{Phys. Lett.}
  {\bfseries B763} (2016) 365--369},
  [\href{https://arxiv.org/abs/1606.04724}{{\ttfamily 1606.04724}}].

\bibitem{Bahjat-Abbas:2017htu}
N.~Bahjat-Abbas, A.~Luna and C.~D. White, \emph{{The Kerr-Schild double copy in
  curved spacetime}},
  \href{http://dx.doi.org/10.1007/JHEP12(2017)004}{\emph{JHEP} {\bfseries 12}
  (2017) 004}, [\href{https://arxiv.org/abs/1710.01953}{{\ttfamily
  1710.01953}}].

\bibitem{Carrillo-Gonzalez:2017iyj}
M.~Carrillo-Gonzalez, R.~Penco and M.~Trodden, \emph{{The classical double copy
  in maximally symmetric spacetimes}},
  \href{http://dx.doi.org/10.1007/JHEP04(2018)028}{\emph{JHEP} {\bfseries 04}
  (2018) 028}, [\href{https://arxiv.org/abs/1711.01296}{{\ttfamily
  1711.01296}}].

\bibitem{DeSmet:2017rve}
P.-J. De~Smet and C.~D. White, \emph{{Extended solutions for the biadjoint
  scalar field}},
  \href{http://dx.doi.org/10.1016/j.physletb.2017.11.007}{\emph{Phys. Lett.}
  {\bfseries B775} (2017) 163--167},
  [\href{https://arxiv.org/abs/1708.01103}{{\ttfamily 1708.01103}}].

\bibitem{Berman:2018hwd}
D.~S. Berman, E.~Chac{\' o}n, A.~Luna and C.~D. White, \emph{{The self-dual
  classical double copy, and the Eguchi-Hanson instanton}},
  \href{https://arxiv.org/abs/1809.04063}{{\ttfamily 1809.04063}}.

\bibitem{Gurses:2018ckx}
M.~Gurses and B.~Tekin, \emph{{Classical Double Copy: Kerr-Schild-Kundt metrics
  from Yang-Mills Theory}},  \href{https://arxiv.org/abs/1810.03411}{{\ttfamily
  1810.03411}}.

\bibitem{Luna:2018dpt}
A.~Luna, R.~Monteiro, I.~Nicholson and D.~O'Connell, \emph{{Type D Spacetimes
  and the Weyl Double Copy}},
  \href{http://dx.doi.org/10.1088/1361-6382/ab03e6}{\emph{Class. Quant. Grav.}
  {\bfseries 36} (2019) 065003},
  [\href{https://arxiv.org/abs/1810.08183}{{\ttfamily 1810.08183}}].

\bibitem{Bahjat-Abbas:2018vgo}
N.~Bahjat-Abbas, R.~Stark-Muchao and C.~D. White, \emph{{Biadjoint wires}},
  \href{http://dx.doi.org/10.1016/j.physletb.2018.11.026}{\emph{Phys. Lett.}
  {\bfseries B788} (2019) 274--279},
  [\href{https://arxiv.org/abs/1810.08118}{{\ttfamily 1810.08118}}].

\bibitem{Andrzejewski:2019hub}
K.~Andrzejewski and S.~Prencel, \emph{{From polarized gravitational waves to
  analytically solvable electromagnetic beams}},
  \href{http://dx.doi.org/10.1103/PhysRevD.100.045006}{\emph{Phys. Rev.}
  {\bfseries D100} (2019) 045006},
  [\href{https://arxiv.org/abs/1901.05255}{{\ttfamily 1901.05255}}].

\bibitem{CarrilloGonzalez:2019gof}
M.~Carrillo~Gonzalez, B.~Melcher, K.~Ratliff, S.~Watson and C.~D. White,
  \emph{{The classical double copy in three spacetime dimensions}},
  \href{http://dx.doi.org/10.1007/JHEP07(2019)167}{\emph{JHEP} {\bfseries 07}
  (2019) 167}, [\href{https://arxiv.org/abs/1904.11001}{{\ttfamily
  1904.11001}}].

\bibitem{Bah:2019sda}
I.~Bah, R.~Dempsey and P.~Weck, \emph{{Kerr-Schild Double Copy and Complex
  Worldlines}},  \href{https://arxiv.org/abs/1910.04197}{{\ttfamily
  1910.04197}}.

\bibitem{Alawadhi:2019urr}
R.~Alawadhi, D.~S. Berman, B.~Spence and D.~Peinador~Veiga, \emph{{S-duality
  and the Double Copy}},  \href{https://arxiv.org/abs/1911.06797}{{\ttfamily
  1911.06797}}.

\bibitem{Banerjee:2019saj}
A.~Banerjee, E.~Colgáin, J.~A. Rosabal and H.~Yavartanoo, \emph{{Ehlers as EM
  duality in the double copy}},
  \href{https://arxiv.org/abs/1912.02597}{{\ttfamily 1912.02597}}.

\bibitem{Lee:2018gxc}
K.~Lee, \emph{{Kerr-Schild Double Field Theory and Classical Double Copy}},
  \href{http://dx.doi.org/10.1007/JHEP10(2018)027}{\emph{JHEP} {\bfseries 10}
  (2018) 027}, [\href{https://arxiv.org/abs/1807.08443}{{\ttfamily
  1807.08443}}].

\bibitem{Siegel:1993xq}
W.~Siegel, \emph{{Two vierbein formalism for string inspired axionic gravity}},
  \href{http://dx.doi.org/10.1103/PhysRevD.47.5453}{\emph{Phys. Rev.}
  {\bfseries D47} (1993) 5453--5459},
  [\href{https://arxiv.org/abs/hep-th/9302036}{{\ttfamily hep-th/9302036}}].

\bibitem{Siegel:1993th}
W.~Siegel, \emph{{Superspace duality in low-energy superstrings}},
  \href{http://dx.doi.org/10.1103/PhysRevD.48.2826}{\emph{Phys. Rev.}
  {\bfseries D48} (1993) 2826--2837},
  [\href{https://arxiv.org/abs/hep-th/9305073}{{\ttfamily hep-th/9305073}}].

\bibitem{Hull:2009mi}
C.~Hull and B.~Zwiebach, \emph{{Double Field Theory}},
  \href{http://dx.doi.org/10.1088/1126-6708/2009/09/099}{\emph{JHEP} {\bfseries
  09} (2009) 099}, [\href{https://arxiv.org/abs/0904.4664}{{\ttfamily
  0904.4664}}].

\bibitem{Hull:2009zb}
C.~Hull and B.~Zwiebach, \emph{{The Gauge algebra of double field theory and
  Courant brackets}},
  \href{http://dx.doi.org/10.1088/1126-6708/2009/09/090}{\emph{JHEP} {\bfseries
  09} (2009) 090}, [\href{https://arxiv.org/abs/0908.1792}{{\ttfamily
  0908.1792}}].

\bibitem{Hohm:2010jy}
O.~Hohm, C.~Hull and B.~Zwiebach, \emph{{Background independent action for
  double field theory}},
  \href{http://dx.doi.org/10.1007/JHEP07(2010)016}{\emph{JHEP} {\bfseries 07}
  (2010) 016}, [\href{https://arxiv.org/abs/1003.5027}{{\ttfamily 1003.5027}}].

\bibitem{Hohm:2010pp}
O.~Hohm, C.~Hull and B.~Zwiebach, \emph{{Generalized metric formulation of
  double field theory}},
  \href{http://dx.doi.org/10.1007/JHEP08(2010)008}{\emph{JHEP} {\bfseries 08}
  (2010) 008}, [\href{https://arxiv.org/abs/1006.4823}{{\ttfamily 1006.4823}}].

\bibitem{Cho:2019ype}
W.~Cho and K.~Lee, \emph{{Heterotic Kerr-Schild Double Field Theory and
  Classical Double Copy}},
  \href{http://dx.doi.org/10.1007/JHEP07(2019)030}{\emph{JHEP} {\bfseries 07}
  (2019) 030}, [\href{https://arxiv.org/abs/1904.11650}{{\ttfamily
  1904.11650}}].

\bibitem{Hohm:2011dz}
O.~Hohm, \emph{{On factorizations in perturbative quantum gravity}},
  \href{http://dx.doi.org/10.1007/JHEP04(2011)103}{\emph{JHEP} {\bfseries 04}
  (2011) 103}, [\href{https://arxiv.org/abs/1103.0032}{{\ttfamily 1103.0032}}].

\bibitem{Cheung:2016say}
C.~Cheung and G.~N. Remmen, \emph{{Twofold Symmetries of the Pure Gravity
  Action}}, \href{http://dx.doi.org/10.1007/JHEP01(2017)104}{\emph{JHEP}
  {\bfseries 01} (2017) 104},
  [\href{https://arxiv.org/abs/1612.03927}{{\ttfamily 1612.03927}}].

\bibitem{Goldberger:2016iau}
W.~D. Goldberger and A.~K. Ridgway, \emph{{Radiation and the classical double
  copy for color charges}},
  \href{http://dx.doi.org/10.1103/PhysRevD.95.125010}{\emph{Phys. Rev.}
  {\bfseries D95} (2017) 125010},
  [\href{https://arxiv.org/abs/1611.03493}{{\ttfamily 1611.03493}}].

\bibitem{Luna:2016hge}
A.~Luna, R.~Monteiro, I.~Nicholson, A.~Ochirov, D.~O'Connell, N.~Westerberg
  et~al., \emph{{Perturbative spacetimes from Yang-Mills theory}},
  \href{http://dx.doi.org/10.1007/JHEP04(2017)069}{\emph{JHEP} {\bfseries 04}
  (2017) 069}, [\href{https://arxiv.org/abs/1611.07508}{{\ttfamily
  1611.07508}}].

\bibitem{Janis:1968zz}
A.~I. Janis, E.~T. Newman and J.~Winicour, \emph{{Reality of the Schwarzschild
  Singularity}},
  \href{http://dx.doi.org/10.1103/PhysRevLett.20.878}{\emph{Phys. Rev. Lett.}
  {\bfseries 20} (1968) 878--880}.

\bibitem{Saotome:2012vy}
R.~Saotome and R.~Akhoury, \emph{{Relationship Between Gravity and Gauge
  Scattering in the High Energy Limit}},
  \href{http://dx.doi.org/10.1007/JHEP01(2013)123}{\emph{JHEP} {\bfseries 01}
  (2013) 123}, [\href{https://arxiv.org/abs/1210.8111}{{\ttfamily 1210.8111}}].

\bibitem{Neill:2013wsa}
D.~Neill and I.~Z. Rothstein, \emph{{Classical Space-Times from the S Matrix}},
  \href{http://dx.doi.org/10.1016/j.nuclphysb.2013.09.007}{\emph{Nucl. Phys.}
  {\bfseries B877} (2013) 177--189},
  [\href{https://arxiv.org/abs/1304.7263}{{\ttfamily 1304.7263}}].

\bibitem{Monteiro:2011pc}
R.~Monteiro and D.~O'Connell, \emph{{The Kinematic Algebra From the Self-Dual
  Sector}}, \href{http://dx.doi.org/10.1007/JHEP07(2011)007}{\emph{JHEP}
  {\bfseries 07} (2011) 007},
  [\href{https://arxiv.org/abs/1105.2565}{{\ttfamily 1105.2565}}].

\bibitem{Anastasiou:2014qba}
A.~Anastasiou, L.~Borsten, M.~J. Duff, L.~J. Hughes and S.~Nagy,
  \emph{{Yang-Mills origin of gravitational symmetries}},
  \href{http://dx.doi.org/10.1103/PhysRevLett.113.231606}{\emph{Phys. Rev.
  Lett.} {\bfseries 113} (2014) 231606},
  [\href{https://arxiv.org/abs/1408.4434}{{\ttfamily 1408.4434}}].

\bibitem{Cardoso:2016ngt}
G.~L. Cardoso, S.~Nagy and S.~Nampuri, \emph{{A double copy for $ \mathcal{N}=2
  $ supergravity: a linearised tale told on-shell}},
  \href{http://dx.doi.org/10.1007/JHEP10(2016)127}{\emph{JHEP} {\bfseries 10}
  (2016) 127}, [\href{https://arxiv.org/abs/1609.05022}{{\ttfamily
  1609.05022}}].

\bibitem{Cardoso:2016amd}
G.~Cardoso, S.~Nagy and S.~Nampuri, \emph{{Multi-centered $ \mathcal{N}=2 $ BPS
  black holes: a double copy description}},
  \href{http://dx.doi.org/10.1007/JHEP04(2017)037}{\emph{JHEP} {\bfseries 04}
  (2017) 037}, [\href{https://arxiv.org/abs/1611.04409}{{\ttfamily
  1611.04409}}].

\bibitem{Anastasiou:2018rdx}
A.~Anastasiou, L.~Borsten, M.~J. Duff, S.~Nagy and M.~Zoccali, \emph{{Gravity
  as Gauge Theory Squared: A Ghost Story}},
  \href{http://dx.doi.org/10.1103/PhysRevLett.121.211601}{\emph{Phys. Rev.
  Lett.} {\bfseries 121} (2018) 211601},
  [\href{https://arxiv.org/abs/1807.02486}{{\ttfamily 1807.02486}}].

\bibitem{Borsten:2019prq}
L.~Borsten, I.~Jubb, V.~Makwana and S.~Nagy, \emph{{Gauge $\times$ Gauge on
  Spheres}},  \href{https://arxiv.org/abs/1911.12324}{{\ttfamily 1911.12324}}.

\bibitem{Goldberger:2017frp}
W.~D. Goldberger, S.~G. Prabhu and J.~O. Thompson, \emph{{Classical gluon and
  graviton radiation from the bi-adjoint scalar double copy}},
  \href{http://dx.doi.org/10.1103/PhysRevD.96.065009}{\emph{Phys. Rev.}
  {\bfseries D96} (2017) 065009},
  [\href{https://arxiv.org/abs/1705.09263}{{\ttfamily 1705.09263}}].

\bibitem{Goldberger:2017vcg}
W.~D. Goldberger and A.~K. Ridgway, \emph{{Bound states and the classical
  double copy}},
  \href{http://dx.doi.org/10.1103/PhysRevD.97.085019}{\emph{Phys. Rev.}
  {\bfseries D97} (2018) 085019},
  [\href{https://arxiv.org/abs/1711.09493}{{\ttfamily 1711.09493}}].

\bibitem{Chester:2017vcz}
D.~Chester, \emph{{Radiative double copy for Einstein-Yang-Mills theory}},
  \href{http://dx.doi.org/10.1103/PhysRevD.97.084025}{\emph{Phys. Rev.}
  {\bfseries D97} (2018) 084025},
  [\href{https://arxiv.org/abs/1712.08684}{{\ttfamily 1712.08684}}].

\bibitem{Goldberger:2017ogt}
W.~D. Goldberger, J.~Li and S.~G. Prabhu, \emph{{Spinning particles, axion
  radiation, and the classical double copy}},
  \href{http://dx.doi.org/10.1103/PhysRevD.97.105018}{\emph{Phys. Rev.}
  {\bfseries D97} (2018) 105018},
  [\href{https://arxiv.org/abs/1712.09250}{{\ttfamily 1712.09250}}].

\bibitem{Li:2018qap}
J.~Li and S.~G. Prabhu, \emph{{Gravitational radiation from the classical
  spinning double copy}},
  \href{http://dx.doi.org/10.1103/PhysRevD.97.105019}{\emph{Phys. Rev.}
  {\bfseries D97} (2018) 105019},
  [\href{https://arxiv.org/abs/1803.02405}{{\ttfamily 1803.02405}}].

\bibitem{Carrillo-Gonzalez:2018pjk}
M.~Carrillo~Gonzalez, R.~Penco and M.~Trodden, \emph{{Radiation of scalar modes
  and the classical double copy}},
  \href{http://dx.doi.org/10.1007/JHEP11(2018)065}{\emph{JHEP} {\bfseries 11}
  (2018) 065}, [\href{https://arxiv.org/abs/1809.04611}{{\ttfamily
  1809.04611}}].

\bibitem{Shen:2018ebu}
C.-H. Shen, \emph{{Gravitational Radiation from Color-Kinematics Duality}},
  \href{https://arxiv.org/abs/1806.07388}{{\ttfamily 1806.07388}}.

\bibitem{Plefka:2018dpa}
J.~Plefka, J.~Steinhoff and W.~Wormsbecher, \emph{{Effective action of dilaton
  gravity as the classical double copy of Yang-Mills theory}},
  \href{https://arxiv.org/abs/1807.09859}{{\ttfamily 1807.09859}}.

\bibitem{Plefka:2019hmz}
J.~Plefka, C.~Shi, J.~Steinhoff and T.~Wang, \emph{{Breakdown of the classical
  double copy for the effective action of dilaton-gravity at NNLO}},
  \href{http://dx.doi.org/10.1103/PhysRevD.100.086006}{\emph{Phys. Rev.}
  {\bfseries D100} (2019) 086006},
  [\href{https://arxiv.org/abs/1906.05875}{{\ttfamily 1906.05875}}].

\bibitem{Goldberger:2019xef}
W.~D. Goldberger and J.~Li, \emph{{Strings, extended objects, and the classical
  double copy}},  \href{https://arxiv.org/abs/1912.01650}{{\ttfamily
  1912.01650}}.

\bibitem{Adamo:2017nia}
T.~Adamo, E.~Casali, L.~Mason and S.~Nekovar, \emph{{Scattering on plane waves
  and the double copy}},
  \href{http://dx.doi.org/10.1088/1361-6382/aa9961}{\emph{Class. Quant. Grav.}
  {\bfseries 35} (2018) 015004},
  [\href{https://arxiv.org/abs/1706.08925}{{\ttfamily 1706.08925}}].

\bibitem{Adamo:2018mpq}
T.~Adamo, E.~Casali, L.~Mason and S.~Nekovar, \emph{{Plane wave backgrounds and
  colour-kinematics duality}},
  \href{https://arxiv.org/abs/1810.05115}{{\ttfamily 1810.05115}}.

\bibitem{Bjerrum-Bohr:2014zsa}
N.~E.~J. Bjerrum-Bohr, J.~F. Donoghue, B.~R. Holstein, L.~Plante and
  P.~Vanhove, \emph{{Bending of Light in Quantum Gravity}},
  \href{http://dx.doi.org/10.1103/PhysRevLett.114.061301}{\emph{Phys. Rev.
  Lett.} {\bfseries 114} (2015) 061301},
  [\href{https://arxiv.org/abs/1410.7590}{{\ttfamily 1410.7590}}].

\bibitem{Bjerrum-Bohr:2016hpa}
N.~E.~J. Bjerrum-Bohr, J.~F. Donoghue, B.~R. Holstein, L.~Plante and
  P.~Vanhove, \emph{{Light-like Scattering in Quantum Gravity}},
  \href{http://dx.doi.org/10.1007/JHEP11(2016)117}{\emph{JHEP} {\bfseries 11}
  (2016) 117}, [\href{https://arxiv.org/abs/1609.07477}{{\ttfamily
  1609.07477}}].

\bibitem{Luna:2017dtq}
A.~Luna, I.~Nicholson, D.~O'Connell and C.~D. White, \emph{{Inelastic Black
  Hole Scattering from Charged Scalar Amplitudes}},
  \href{http://dx.doi.org/10.1007/JHEP03(2018)044}{\emph{JHEP} {\bfseries 03}
  (2018) 044}, [\href{https://arxiv.org/abs/1711.03901}{{\ttfamily
  1711.03901}}].

\bibitem{Kosower:2018adc}
D.~A. Kosower, B.~Maybee and D.~O'Connell, \emph{{Amplitudes, Observables, and
  Classical Scattering}},
  \href{http://dx.doi.org/10.1007/JHEP02(2019)137}{\emph{JHEP} {\bfseries 02}
  (2019) 137}, [\href{https://arxiv.org/abs/1811.10950}{{\ttfamily
  1811.10950}}].

\bibitem{Chung:2018kqs}
M.-Z. Chung, Y.-T. Huang, J.-W. Kim and S.~Lee, \emph{{The simplest massive
  S-matrix: from minimal coupling to Black Holes}},
  \href{http://dx.doi.org/10.1007/JHEP04(2019)156}{\emph{JHEP} {\bfseries 04}
  (2019) 156}, [\href{https://arxiv.org/abs/1812.08752}{{\ttfamily
  1812.08752}}].

\bibitem{Bern:2019nnu}
Z.~Bern, C.~Cheung, R.~Roiban, C.-H. Shen, M.~P. Solon and M.~Zeng,
  \emph{{Scattering Amplitudes and the Conservative Hamiltonian for Binary
  Systems at Third Post-Minkowskian Order}},
  \href{http://dx.doi.org/10.1103/PhysRevLett.122.201603}{\emph{Phys. Rev.
  Lett.} {\bfseries 122} (2019) 201603},
  [\href{https://arxiv.org/abs/1901.04424}{{\ttfamily 1901.04424}}].

\bibitem{Bautista:2019tdr}
Y.~F. Bautista and A.~Guevara, \emph{{From Scattering Amplitudes to Classical
  Physics: Universality, Double Copy and Soft Theorems}},
  \href{https://arxiv.org/abs/1903.12419}{{\ttfamily 1903.12419}}.

\bibitem{Maybee:2019jus}
B.~Maybee, D.~O'Connell and J.~Vines, \emph{{Observables and amplitudes for
  spinning particles and black holes}},
  \href{https://arxiv.org/abs/1906.09260}{{\ttfamily 1906.09260}}.

\bibitem{Guevara:2019fsj}
A.~Guevara, A.~Ochirov and J.~Vines, \emph{{Black-hole scattering with general
  spin directions from minimal-coupling amplitudes}},
  \href{http://dx.doi.org/10.1103/PhysRevD.100.104024}{\emph{Phys. Rev.}
  {\bfseries D100} (2019) 104024},
  [\href{https://arxiv.org/abs/1906.10071}{{\ttfamily 1906.10071}}].

\bibitem{Arkani-Hamed:2019ymq}
N.~Arkani-Hamed, Y.-t. Huang and D.~O'Connell, \emph{{Kerr Black Holes as
  Elementary Particles}},  \href{https://arxiv.org/abs/1906.10100}{{\ttfamily
  1906.10100}}.

\bibitem{Johansson:2019dnu}
H.~Johansson and A.~Ochirov, \emph{{Double copy for massive quantum particles
  with spin}}, \href{http://dx.doi.org/10.1007/JHEP09(2019)040}{\emph{JHEP}
  {\bfseries 09} (2019) 040},
  [\href{https://arxiv.org/abs/1906.12292}{{\ttfamily 1906.12292}}].

\bibitem{Huang:2019cja}
Y.-t. Huang, U.~Kol and D.~O'Connell, \emph{{The Double Copy of
  Electric-Magnetic Duality}},
  \href{https://arxiv.org/abs/1911.06318}{{\ttfamily 1911.06318}}.

\bibitem{Bern:2019crd}
Z.~Bern, C.~Cheung, R.~Roiban, C.-H. Shen, M.~P. Solon and M.~Zeng,
  \emph{{Black Hole Binary Dynamics from the Double Copy and Effective
  Theory}}, \href{http://dx.doi.org/10.1007/JHEP10(2019)206}{\emph{JHEP}
  {\bfseries 10} (2019) 206},
  [\href{https://arxiv.org/abs/1908.01493}{{\ttfamily 1908.01493}}].

\bibitem{Bautista:2019evw}
Y.~F. Bautista and A.~Guevara, \emph{{On the Double Copy for Spinning Matter}},
   \href{https://arxiv.org/abs/1908.11349}{{\ttfamily 1908.11349}}.

\bibitem{Moynihan:2019bor}
N.~Moynihan, \emph{{Kerr-Newman from Minimal Coupling}},
  \href{https://arxiv.org/abs/1909.05217}{{\ttfamily 1909.05217}}.

\bibitem{Kalin:2019rwq}
G.~Kalin and R.~A. Porto, \emph{{From Boundary Data to Bound States}},
  \href{https://arxiv.org/abs/1910.03008}{{\ttfamily 1910.03008}}.

\bibitem{Plefka:2019wyg}
J.~Plefka, C.~Shi and T.~Wang, \emph{{The Double Copy of Massive Scalar-QCD}},
  \href{https://arxiv.org/abs/1911.06785}{{\ttfamily 1911.06785}}.

\bibitem{Godazgar:2019ikr}
H.~Godazgar, M.~Godazgar and C.~N. Pope, \emph{{Taub-NUT from the Dirac
  monopole}},
  \href{http://dx.doi.org/10.1016/j.physletb.2019.134938}{\emph{Phys. Lett.}
  {\bfseries B798} (2019) 134938},
  [\href{https://arxiv.org/abs/1908.05962}{{\ttfamily 1908.05962}}].

\bibitem{Bern:2010yg}
Z.~Bern, T.~Dennen, Y.-t. Huang and M.~Kiermaier, \emph{{Gravity as the Square
  of Gauge Theory}},
  \href{http://dx.doi.org/10.1103/PhysRevD.82.065003}{\emph{Phys. Rev.}
  {\bfseries D82} (2010) 065003},
  [\href{https://arxiv.org/abs/1004.0693}{{\ttfamily 1004.0693}}].

\bibitem{Tolotti:2013caa}
M.~Tolotti and S.~Weinzierl, \emph{{Construction of an effective Yang-Mills
  Lagrangian with manifest BCJ duality}},
  \href{http://dx.doi.org/10.1007/JHEP07(2013)111}{\emph{JHEP} {\bfseries 07}
  (2013) 111}, [\href{https://arxiv.org/abs/1306.2975}{{\ttfamily 1306.2975}}].

\bibitem{Ett:2010by}
B.~Ett and D.~Kastor, \emph{{An Extended Kerr-Schild Ansatz}},
  \href{http://dx.doi.org/10.1088/0264-9381/27/18/185024}{\emph{Class. Quant.
  Grav.} {\bfseries 27} (2010) 185024},
  [\href{https://arxiv.org/abs/1002.4378}{{\ttfamily 1002.4378}}].

\bibitem{Burgess:1994kq}
C.~P. Burgess, R.~C. Myers and F.~Quevedo, \emph{{On spherically symmetric
  string solutions in four-dimensions}},
  \href{http://dx.doi.org/10.1016/S0550-3213(95)00090-9}{\emph{Nucl. Phys.}
  {\bfseries B442} (1995) 75--96},
  [\href{https://arxiv.org/abs/hep-th/9410142}{{\ttfamily hep-th/9410142}}].

\bibitem{Monteiro:2018xev}
R.~Monteiro, I.~Nicholson and D.~O'Connell, \emph{{Spinor-helicity and the
  algebraic classification of higher-dimensional spacetimes}},
  \href{https://arxiv.org/abs/1809.03906}{{\ttfamily 1809.03906}}.

\bibitem{Jeon:2011cn}
I.~Jeon, K.~Lee and J.-H. Park, \emph{{Stringy differential geometry, beyond
  Riemann}}, \href{http://dx.doi.org/10.1103/PhysRevD.84.044022}{\emph{Phys.
  Rev.} {\bfseries D84} (2011) 044022},
  [\href{https://arxiv.org/abs/1105.6294}{{\ttfamily 1105.6294}}].

\bibitem{Jeon:2010rw}
I.~Jeon, K.~Lee and J.-H. Park, \emph{{Differential geometry with a projection:
  Application to double field theory}},
  \href{http://dx.doi.org/10.1007/JHEP04(2011)014}{\emph{JHEP} {\bfseries 04}
  (2011) 014}, [\href{https://arxiv.org/abs/1011.1324}{{\ttfamily 1011.1324}}].

\bibitem{Park:2015bza}
J.-H. Park, S.-J. Rey, W.~Rim and Y.~Sakatani, \emph{{O(D, D) covariant Noether
  currents and global charges in double field theory}},
  \href{http://dx.doi.org/10.1007/JHEP11(2015)131}{\emph{JHEP} {\bfseries 11}
  (2015) 131}, [\href{https://arxiv.org/abs/1507.07545}{{\ttfamily
  1507.07545}}].

\bibitem{Angus:2018mep}
S.~Angus, K.~Cho and J.-H. Park, \emph{{Einstein Double Field Equations}},
  \href{http://dx.doi.org/10.1140/epjc/s10052-018-5982-y}{\emph{Eur. Phys. J.}
  {\bfseries C78} (2018) 500},
  [\href{https://arxiv.org/abs/1804.00964}{{\ttfamily 1804.00964}}].

\end{thebibliography}\endgroup
\bibliographystyle{JHEP}

\end{document}